\documentclass[tradiabstract,longauth]{aa} 

\usepackage{graphicx}
\usepackage{rotating}
\usepackage{numprint}
\usepackage{lscape}
\usepackage{txfonts}
\usepackage{natbib}
\usepackage{longtable}
\usepackage{multicol}
\usepackage{hyperref}
\hypersetup{breaklinks=true}
\usepackage{color}
\usepackage{gensymb}

\newcommand{\CIV}{{\rm C~{\sc iv}}}

\newcommand{\MgII}{{\rm Mg~{\sc ii}}}

\newcommand{\kms}{{\rm km}\,{\rm s}^{-1}}

\newcommand{\lya}{Lyman-$\alpha$}

\def\Sec#1{Section~\ref{s:#1}}

\def\Fig#1{Fig.~\ref{fig:#1}}
\def\Tab#1{Table~\ref{t:#1}}
\def\ion#1#2{{\rm #1~{\sc #2}}}

\begin{document}

   \title{The Sloan Digital Sky Survey Quasar Catalog: Fourteenth Data Release\thanks{http://www.sdss.org/dr14/algorithms/qso\_catalog}}

   \author{Isabelle P\^aris
          \inst{1}
          \and
          Patrick Petitjean
          \inst{2}
          \and
          \'Eric Aubourg
          \inst{3}
          \and
          Adam D. Myers
          \inst{4}
          \and
          Alina Streblyanska
          \inst{5,6}
          \and
          Brad W. Lyke
          \inst{4}
          \and
          Scott F. Anderson
          \inst{7}
          \and
          \'Eric Armengaud
          \inst{8}
          \and
          Julian Bautista
          \inst{9}
          \and
          Michael R. Blanton
          \inst{10}
          \and
          Michael Blomqvist
          \inst{1}
          \and
          Jonathan Brinkmann
          \inst{11}
          \and
          Joel R. Brownstein
          \inst{9}
          \and
          William Nielsen Brandt
          \inst{12,13,14}
          \and
          \'Etienne Burtin
          \inst{8}
          \and
          Kyle Dawson
          \inst{9}
          \and
          Sylvain de la Torre
          \inst{1}
          \and
          Antonis Georgakakis
          \inst{15}
          \and
          H\'ector Gil-Mar\'in
          \inst{16,17}
          \and
          Paul J. Green
          \inst{18}
          \and
          Patrick B. Hall
          \inst{19}
          \and
          Jean-Paul Kneib
          \inst{20}
          \and
          Stephanie M. LaMassa
          \inst{21}
          \and
          Jean-Marc Le Goff
          \inst{8}
          \and
          Chelsea MacLeod
          \inst{18}
          \and
          Vivek Mariappan
          \inst{9}
          \and
          Ian D. McGreer
          \inst{22}
          \and
          Andrea Merloni
          \inst{15}
          \and
          Pasquier Noterdaeme
          \inst{2}
          \and
          Nathalie Palanque-Delabrouille
          \inst{8}
          \and
          Will J. Percival
          \inst{23}
          \and
          Ashley J. Ross
          \inst{24}
          \and
          Graziano Rossi
          \inst{25}
          \and
          Donald P. Schneider
          \inst{12,13}
          \and
          Hee-Jong Seo
          \inst{26}
          \and
          Rita Tojeiro
          \inst{27}
          \and
          Benjamin A. Weaver
          \inst{28}
          \and
          Anne-Marie Weijmans
          \inst{27}
          \and
          Christophe Y\`eche
          \inst{8}
          \and
          Pauline Zarrouk
          \inst{8}
          \and
          Gong-Bo Zhao
          \inst{29,23}
          }

   \institute{
           Aix Marseille Univ, CNRS, LAM, Laboratoire d’Astrophysique de Marseille, Marseille, France\\
              \email{isabelle.paris@lam.fr}
         \and
           Institut d‘Astropysique de Paris, UMR 7095, CNRS-UPMC, 98bis bd Arago, 75014 Paris, France\\
            	\email{petitjean@iap.fr}
         \and
           APC, University of Paris Diderot, CNRS/IN2P3, CEA/IRFU, Observatoire de Paris, Sorbonne Paris Cite, France
         \and
           Department of Physics and Astronomy, University of Wyoming, Laramie, WY 82071, USA
         \and
           Instituto de Astrof\'isica de Canarias, E-38205 La Laguna, Tenerife, Spain
         \and
          Departamento de Astrof\'isica, Universidad de La Laguna (ULL), E-38206 La Laguna, Tenerife, Spain
         \and
           Department of Astronomy, Box 351580, University of Washington, Seattle, WA 98195, USA
         \and
            CEA, Centre de Saclay, IRFU, F-91191, Gif-sur-Yvette, France
         \and
           Department of Physics and Astronomy, University of Utah, 115 S. 1400 E., Salt Lake City, UT 84112, USA
         \and
           Center for Cosmology and Particle Physics, Department of Physics, New York University, 726 Broadway, Room 1005, New York, NY 10003, USA
          \and
           Apache Point Observatory, P.O. Box 59, Sunspot, NM 88349, USA
          \and
             Department of Astronomy and Astrophysics, Eberly College of Science, The Pennsylvania State University, 525 Davey Laboratory, University Park, PA 16802, USA
          \and
			Institute for Gravitation and the Cosmos, The Pennsylvania State University, University Park, PA 16802, USA   
		  \and
		    Department of Physics, The Pennsylvania State University, University Park, PA 16802, USA
		  \and
		    Max-Planck-Institut f\"ur Extraterrestrische Physik, Gie\ss enbachstr. 1, D-85748 Garching, Germany
		  \and
		    Sorbonne Universit\'es, Institut Lagrange de Paris (ILP), 98 bis Boulevard Arago, 75014 Paris, France
		  \and
		     Laboratoire de Physique Nucl\'eaire et de Hautes Energies, Universit\'e Pierre et Marie Curie, 4 Place Jussieu, 75005 Paris, France
		  \and
		    Harvard-Smithsonian Center for Astrophysics, 60 Garden St., Cambridge, MA 02138, USA
		  \and
            Department of Physics and Astronomy, York University, 4700 Keele St., Toronto, ON, M3J 1P3, Canada
          \and
		    Institute of Physics, Laboratory of Astrophysics, Ecole Polytechnique F\'ed\'erale de Lausanne (EPFL), Observatoire de Sauverny, 1290 Versoix, Switzerland
		  \and
		    Space Telescope Science Institute, 3700 San Martin Drive, Baltimore, MD 21218, USA
		  \and
		    Steward Observatory, The University of Arizona, 933 North Cherry Avenue, Tucson, AZ 85721–0065, USA
		  \and
		    Institute of Cosmology\& Gravitation, University of Portsmouth, Dennis Sciama Building, Portsmouth, PO1 3FX, UK
		  \and
		     Center for Cosmology and AstroParticle Physics, The Ohio State University, 191 W. Woodruff Ave., Columbus, OH 43210, USA
		  \and
          Department of Physics and Astronomy, Sejong University, Seoul, 143-747, Korea
          \and
            Department of Physics and Astronomy, Ohio University, Clippinger Labs, Athens, OH 45701
          \and
		     School of Physics and Astronomy, University of St Andrews, North Haugh, St Andrews, KY16 9SS
          \and
            National Optical Astronomy Observatory, 950 North Cherry Avenue, Tucson, AZ 85719, USA
		  \and
		    National Astronomical Observatories, Chinese Academy of Sciences, 20A Datun Road, Chaoyang District, Beijing 100012, China
             }

   \date{Received \today; accepted XXX}

  \abstract{
We present the Data Release 14 Quasar catalog (DR14Q) from the extended Baryon Oscillation Spectroscopic Survey (eBOSS) of the Sloan Digital Sky Survey IV (SDSS-IV). This catalog includes all SDSS-IV/eBOSS objects that were spectroscopically targeted as quasar candidates and  that  are  confirmed  as  quasars  via  
 a new automated procedure combined with a partial visual inspection of spectra,  have  luminosities
$M_{\rm i} \left[ z=2 \right] < -20.5$  (in a $\Lambda$CDM cosmology with $H_0 = 70 \ {\rm km \ s^{-1} \ Mpc ^{-1}}$, $\Omega _{\rm M} = 0.3$, and $\Omega _{\rm \Lambda} = 0.7$), and either display at least one emission line with a full width at half maximum (FWHM) larger than $500 \ {\rm km \ s^{-1}}$ or, if not, have interesting/complex absorption features. 
The catalog also includes previously spectroscopically-confirmed quasars from SDSS-I, II and III. The catalog contains \numprint{526356} quasars (\numprint{144046} are new discoveries since the beginning of SDSS-IV) detected over \numprint{9376} deg$^2$ (\numprint{2044} deg$^2$ having new spectroscopic data available)
 with robust identification and redshift measured by a combination of principal component eigenspectra. The catalog is estimated to have about 0.5\%~ contamination. 
 Redshifts are provided for the \ion{Mg}{ii} emission line. 
 The catalog identifies \numprint{21877} broad absorption line quasars and lists their characteristics. For each object, the catalog presents five-band ($u$, $g$, $r$, $i$, $z$) CCD-based photometry with typical accuracy of 0.03 mag. 
 The catalog also contains X-ray, ultraviolet, near-infrared, and radio emission properties of the quasars, when available, from other large-area surveys. The calibrated digital spectra, covering the wavelength region \numprint{3610}--\numprint{10140} \AA\ at a spectral resolution in the range \numprint{1300}$< R <$ \numprint{2500}, can be retrieved  from the SDSS Science Archiver Server. 
  }

   \keywords{catalogs --
                surveys --
                quasars: general
               }

   \maketitle
%
\section{Introduction}

Since the identification of the first quasar redshift by \cite{schmidt1963}, each generation of spectroscopic surveys has enlarged the 
number of known quasars by roughly an order of magnitude: the Bright Quasar Survey \citep{schmidt1983} reached the 100 discoveries milestone, followed by the Large Bright Quasar Survey \citep[LBQS; ][]{hewett1995} and 
its \numprint{1000} objects, then the $\sim$\numprint{25000} quasars from the 2dF Quasar Redshift Survey 
\citep[2QZ; ][]{croom2004}, and the Sloan Digital Sky Survey \citep[SDSS; ][]{york2000} with 
over \numprint{100000} new quasars \citep{schneider2010}.
Many other surveys have also significantly contributed to increase the number of known quasars \citep[e.g.][]{osmer1980,boyle1988,storrie1996}.
\\

Each iteration of SDSS has pursued different science goals, and hence set different requirements for their associated quasar target selection. 
\\
SDSS-I/II \citep{york2000} aimed to observe $\sim 10^5$ quasars; The final quasar list was presented in the SDSS Data Release 7 (DR7) quasar catalog \citep{schneider2010}.
The main science driver was studies of the quasar population through the measurement of their luminosity function \citep[e.g. ][]{richards2006} and clustering properties \citep[e.g. ][]{hennawi2006,shen2007}.
The quasar program of SDSS-I/II also led to the discovery of a significant sample of $z > 5$ quasars \citep[e.g. ][]{fan2006,jiang2008}, large samples of broad absorption line (BAL) quasars \citep[e.g. ][]{reichard2003a,trump2006,Gibson2008}, type 2 candidates \citep{reyes2008} or samples of objects with peculiar properties such as weak emission lines \citep{Diamond-Stanic2009}.
Quasar target selection algorithms for SDSS-I/II are fully detailed in \cite{richards2002} and \cite{schneider2010}.

The main motivation to observe quasars with SDSS-III/BOSS \citep{eisenstein2011,dawson2013} was 
to constrain the Baryon Acoustic Oscillation (BAO) scale at $z \sim 2.5$ using the \ion{H}{i} located in the intergalactic medium (IGM) as a tracer of large scale structures.
About \numprint{270000} quasars, mostly in the redshift range 2.15--3.5 for which at least part of the
\lya\ forest lies in the spectral range, have been discovered by SDSS-III/BOSS.
The measurement of the auto-correlation function of the \lya\ forest \citep[e.g. ][]{bautista2017} and the cross-correlation of quasars and the \lya\ forest \citep[e.g. ][]{dumasdesbourboux2017} have provided unprecedented cosmological constraints at $z \sim 2.5$.
This sample was also used to study the luminosity function of quasars \citep{ross2013,palanque2013a}, moderate-scale clustering of $z \sim 2.5$ quasars \citep[e.g. ][]{eftekharzadeh2015}.
Repeat spectroscopic observations of BAL quasars have been performed to constrain the scale and dynamics of quasar outflows \citep{filizak2012,filizak2013,filizak2014}.
Peculiar population of quasars have been also identified in this enormous sample such as $z >2$ type 2 quasar candidates \citep{alexandroff2013} or extremely red quasars \citep{ross2015,hamann2017}.
In order to maximize the number of $z > 2$ quasars, the target selection for SDSS-III/BOSS used a variety of target selection algorithms \citep{bovy2011,Bovy2012,kirkpatrick2011,yeche2010,palanque2011,richards2004}.
The overall quasar target selection strategy is described in \cite{ross2012}.
\\

Quasar observation in SDSS-IV is driven by multiple scientific goals such as cosmology, understanding the physical nature of X-ray sources and variable sources.

SDSS-IV/eBOSS aims to constrain the angular-diameter distance $d_A \left( z \right)$ and the Hubble parameter $H \left( z \right)$ in the redshift range 0.6--3.5 using four different tracers of the underlying density field over \numprint{7500} ${\rm deg ^2}$: \numprint{250000} new luminous red galaxies (LRG) at $0.6 < z < 1.0$, \numprint{195000} new emission line galaxies (ELG) at $0.7 < z < 1.1$, \numprint{450000} new quasars at $0.9 < z < 2.2$, and the \lya\ forest of \numprint{60000} new $z > 2.2$ quasars.

SDSS-IV/SPIDERS (SPectroscopic IDentification of ERosita Sources) investigates the nature of X-ray emitting sources, including active galactic nuclei \citep{dwelly2017} and galaxy clusters \citep{clerc2016}.
Initially, SPIDERS targets
X-ray sources detected mainly in the ROSAT All Sky
Survey \citep{voges1999,voges2000} which has recently been reprocessed \citep{boller2016}. In late 2018,
SPIDERS plans to begin targeting sources from the eROSITA instrument on
board the Spectrum Roentgen Gamma satellite \citep{predehl2010,merloni2012}.
About 5\% of eBOSS fibers are allocated to SPIDERS targets.
A total of \numprint{22000} spectra of active galactic nuclei are expected by the end of the survey, about \numprint{5000} of them being also targeted by SDSS-IV/eBOSS.

Finally, SDSS-IV/TDSS (Time Domain Spectroscopic Survey) that aims to characterize the physical nature of time-variable sources, primarily on sources detected to be
variable in Pan-STARRS1 data \cite[PS1; ][]{kaiser2010}
or between SDSS and PS1 imaging, has been allocated about 5\% of eBOSS fibers \citep{morganson2015,macleod2017}. 
The targets identified
in PS1 are a mix of quasars (about 60\%)
and stellar variables (about 40\%).
It will lead to the observation of about \numprint{120000} quasars with a majority of them also targeted by SDSS-IV/eBOSS.

This paper presents the SDSS-IV/eBOSS quasar catalog, denoted DR14Q, that compiles all the spectroscopically-confirmed quasars identified
in the course of any of the SDSS iterations and released as part of the SDSS Fourteenth Data Release \citep{DR14}. 
The bulk of the newly discovered quasars contained in DR14Q arise from the main SDSS-IV/eBOSS quasar target selection \citep{myers2015}. The rest  were observed
by ancillary programs \citep[\numprint{83430} quasars not targeted by the SDSS-IV/eBOSS main quasar survey; see ][]{dawson2013,DR10,DR12}, and 
TDSS and SPIDERS (\numprint{27547} and \numprint{1090}, respectively).

We summarize the target selection and observations in \Sec{observations}. We describe the visual inspection process and describe the definition of the DR14Q parent sample in \Sec{construction}.
We discuss the accuracy of redshift estimates in \Sec{redshift} and present our automated detection of BAL quasars in \Sec{bal}.
General properties of the DR14Q sample are reviewed in \Sec{sample} and \Sec{multilambda}, and the format of the catalog is described in \Sec{description}.
 Finally, we conclude in \Sec{conclusion}.

In the following, we will use a ${\rm \Lambda CDM}$ cosmology with ${\rm H_0 = 70 \ km \ s^{-1} \ Mpc^{-1}}$, ${\rm \Omega _M} = 0.3$, ${\rm \Omega _{\Lambda}} = 0.7$ \citep{spergel2003}.
We define a quasar as an object with a luminosity ${\rm M_i \left[ z = 2 \right] < -20.5}$ and either displaying at least one emission line with ${\rm FWHM > 500 \ km s^{-1}}$ or, if not, having interesting/complex absorption features.
Indeed, a few tens of objects have weak emission lines but the \lya\ forest is clearly visible in their spectra \citep{Diamond-Stanic2009}, and thus they are included in the DR14Q catalog. About \numprint{200} quasars with unusual broad absorption lines are also included in our catalog \citep{hall2002} even though they do not formally meet the requirement on emission-line width.
All magnitudes quoted here are Point Spread Function (PSF) magnitudes \citep{stoughton2002} and are corrected for Galactic extinction \citep{schlafly2011}.

\section{Survey outline}
\label{s:observations}

In this Section, we focus on imaging data used to perform the target selection of SDSS-IV quasar programs and new spectroscopic data obtained since August 2014.

\subsection{Imaging data}
\label{s:imaging}

Three sources of imaging data have been used to target quasars in SDSS-IV/eBOSS \citep[full details can be found in][]{myers2015}: updated calibrations of SDSS imaging, the Wide-Field Infrared Survey \citep[WISE; ][]{wright2010}, and the Palomar Transient Factory \citep[PTF; ][]{rau2009,law2009}.

SDSS  Imaging data were gathered using the 2.5m wide-field Sloan telescope \citep{gunn2006} to collect light for a camera with 30 2k$\times$2k CCDs \citep{gunn1998} over five broad bands - \textit{ugriz} \citep{fukugita1996}.
A total of \numprint{14555} unique square degrees of the sky were imaged by this camera, including contiguous areas of $\sim$\numprint{7500}~${\rm deg^2}$ in the North Galactic Cap (NGC) and $\sim$\numprint{3100} ${\rm deg^2}$ in the SGC that comprise the uniform ``Legacy'' areas of the SDSS \citep{DR8}. These data were acquired on dark photometric nights of good seeing \citep{hogg2001}.  
Objects were detected and their properties were measured by the photometric pipeline \citep{lupton2001,stoughton2002} and calibrated photometrically \citep{smith2002,ivezic2004,tucker2006,padmanabhan2008}, and astrometrically \citep{pier2003}.
Targeting for eBOSS is conducted using SDSS imaging that is calibrated to the \cite{schlafly2012} Pan-STARRS solution \citep{finkbeiner2016}. These imaging data were publicly released as part of SDSS-DR13 \citep{DR13}.

The quasar target selection for SDSS-IV/eBOSS also makes use of the W1 and W2 WISE bands centered on 3.4 and 4.6 $\mu {\rm m}$.
The ``unWISE'' coadded photometry is applied to sources detected in the SDSS imaging data as described in  \cite{lang2014}. 
This approach produces photometry of custom coadds of the WISE imaging at the position of all SDSS primary sources.

Imaging data from PTF is also used to target quasars using variability in SDSS-IV/eBOSS. 
Starting from the individual calibrated frames available from IPAC \citep[Infrared Processing \& Analysis Center; ][]{laher2014}, a customized pipeline is applied to build coadded PTF images on a timescale adapted to quasar targeting, i.e. typically 1--4 epochs per year, depending on the cadence and total exposure time within each field.
A stack of all PTF imaging epochs is also constructed to create a catalog of PTF sources.
Finally, light curves are created using coadded PTF images to perform the selection of quasar candidates.

\subsection{Target Selection}
\label{s:qts}

In order to achieve a precision of 2.8\% on $d_A \left( z \right)$ and 4.2\% on $H \left( z \right)$ measurement with the quasar sample, it is necessary to achieve a surface density of at least 58 quasars with $0.9 < z < 2.2$  per square degree \citep{dawson2016}.
The SDSS-IV/eBOSS ``CORE'' sample is intended to recover sufficient quasars in this specific redshift range  and additional quasars at $z > 2.2$ to supplement SDSS-III/BOSS.
The CORE sample homogeneously targets quasars at all redshifts $z > 0.9$  based on the XDQSOz method \citep{Bovy2012} in the  optical and a WISE-optical color cut.
 To be selected, it is required that point sources have a XDQSOz probability to be a $z > 0.9$ quasar larger than 0.2 \textit{and} pass the color cut 
 $m_{\rm opt} - m_{\rm WISE} \geq \left( g -i \right) + 3$, where $m_{\rm opt}$ is a weighted stacked magnitude in the $g$, $r$ and $i$ bands and $m_{\rm WISE}$ is a weighted stacked
 magnitude in the W1 and W2 bands.
Quasar candidates have $g < 22$ \textit{or} $r < 22$ with a surface density of confirmed new quasars (at any redshifts) of $\sim 70 \ {\rm deg^{-2}}$.

SDSS-IV/eBOSS also selects quasar candidates over a wide range of redshifts using their photometric variability measured from the PTF.
In the following we will refer to this sample as the ``PTF'' sample. 
These targets have $r > 19$ and $g < 22.5$ and provide an additional 3-4 $z > 2.1$ quasars per ${\rm deg^2}$.

In addition, known quasars with low quality SDSS-III/BOSS spectra ($0.75 < {\rm S/N \ per \ pixel} < 3$)\footnote{This measurement refers to the standard SDSS spectroscopic sampling of $69 {\rm km \ s^{-1}}$ per pixel.} or with bad spectra are re-observed.

Finally, quasars within 1\arcsec of a radio detection in the FIRST point source catalog \citep{becker1995} are targeted.

A fully detailed description of the quasar target selection in SDSS-IV/eBOSS and a discussion of its performance can be found in \cite{myers2015}.\\

TDSS targets point sources that are selected to be variable in the $g$, $r$ and $i$ bands using the SDSS-DR9 imaging data \citep{DR9} and the multi-epoch Pan-STARRS1 (PS1) photometry
\citep{kaiser2002,kaiser2010}.
The survey does not specifically target quasars in general but a significant fraction of targets belong
to this class \citep{morganson2015}. Furthermore, there are smaller sub-programs (comprising 10\% of the main TDSS survey) that target quasars specifically \citep{macleod2017}.
Therefore, these quasars are included in the parent sample for the quasar catalog.

Finally, the AGN component of SPIDERS targets X-ray sources detected in the concatenation of the Bright and Faint ROSAT All Sky Survey (RASS) catalogs \citep{voges1999,voges2000} and that have an
optical counterpart detected in the DR9 imaging data \citep{DR9}. Objects with $17 < r < 22$ that lie within 1\arcmin\ of a RASS source are targeted.
Details about the AGN target selection are available in \cite{dwelly2017}.

\subsection{Spectroscopy}

\begin{figure}[htbp]
	\centering{\includegraphics[width=\linewidth]{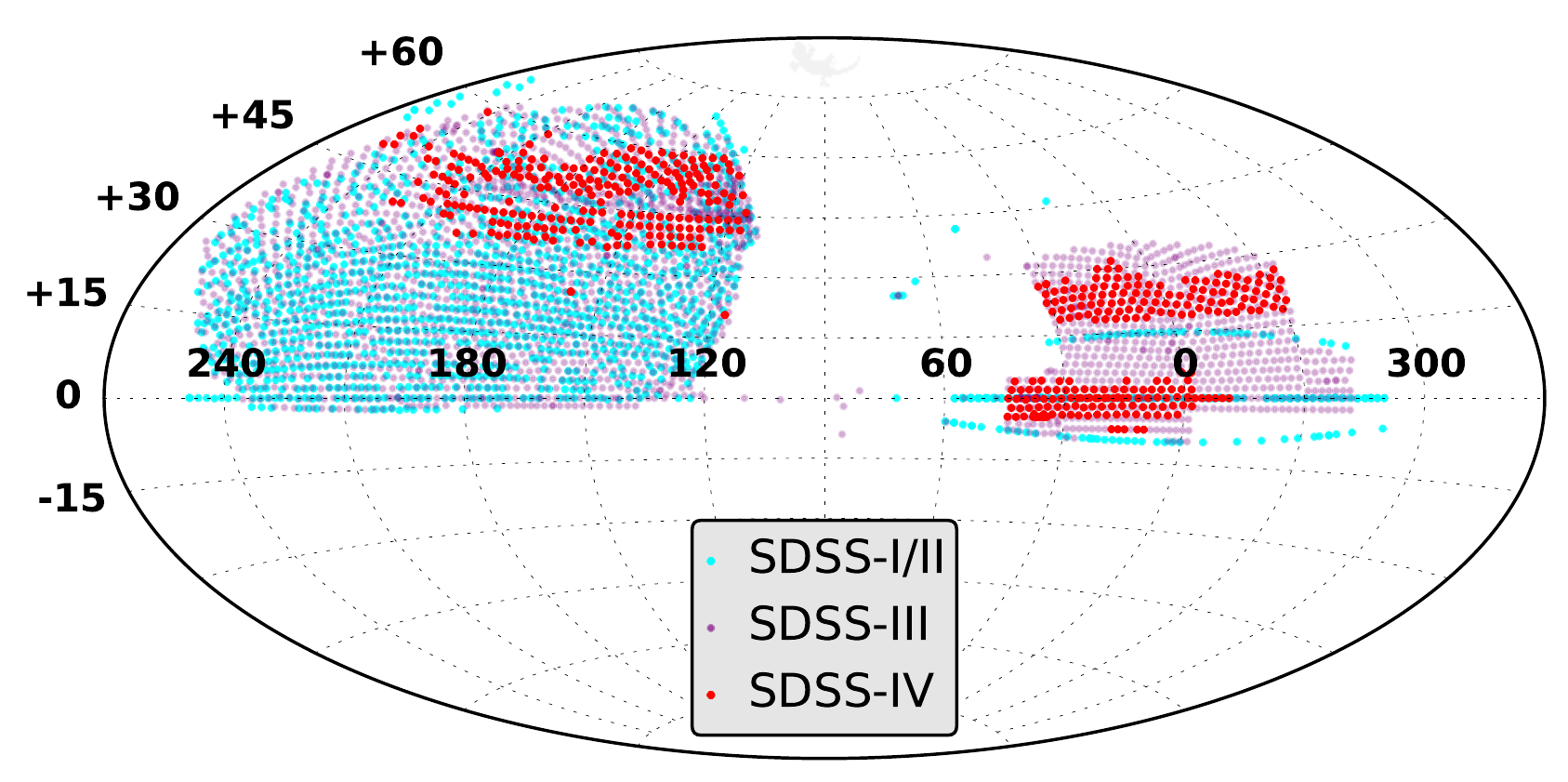}}
\caption{Distribution on the sky of the SDSS-DR14/eBOSS spectroscopy in J2000 equatorial coordinates (expressed in decimal degrees). Cyan dots correspond to the \numprint{1462} plates observed as part of SDSS-I/II. The purple area indicates the \numprint{2587} plates observed as part of SDSS-III/BOSS. The red area represents to the \numprint{496} new plates observed as part of SDSS-IV (i.e. with ${\rm MJD} \geq 56898$). 
}
\label{fig:SkyCoverage}
\end{figure}

\begin{figure}[htbp]
	\centering{\includegraphics[width=\linewidth]{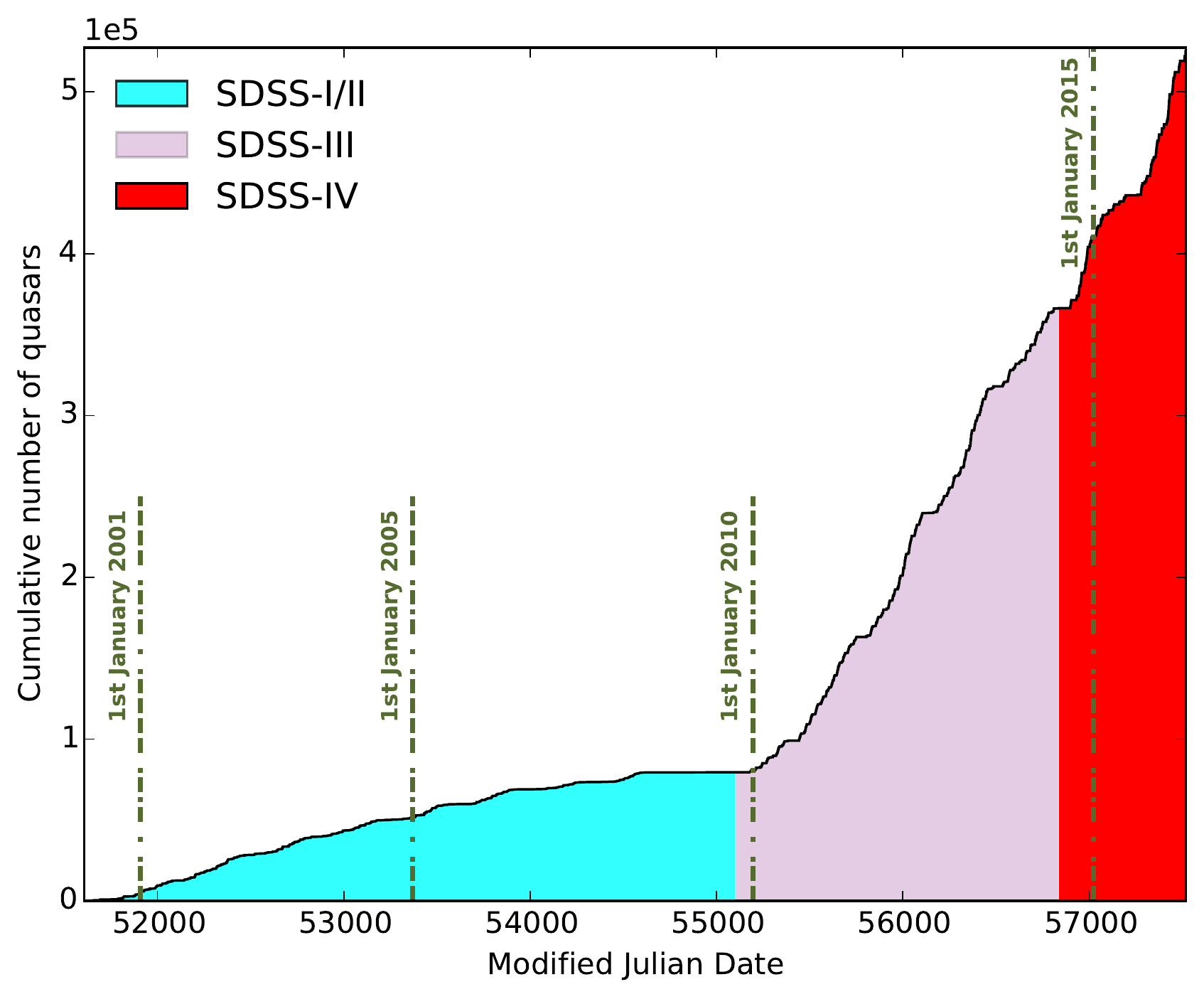}}
\caption{ 
Cumulative number of quasars as a function of observation date during all the iterations of SDSS \citep{york2000,eisenstein2011,blanton2017}. 
Vertical lines show the equivalence between modified Julian dates and usual calendar dates.
A total of \numprint{526356} quasars have been spectroscopically confirmed.
The ``flat portions'' in this figure correspond to annual summer shutdown for the telescope maintenance.
}
\label{fig:ProgressPlot}
\end{figure}

Spectroscopic data for SDSS-IV are acquired in a similar manner as for SDSS-III \citep{dawson2016}.
Targets identified by the various selection algorithms are observed with the BOSS spectrographs whose resolution varies from  
$\sim$\numprint{1300} at \numprint{3600}~\AA~ to \numprint{2500} at \numprint{10000}~\AA~ \citep{smee2013}.
Spectroscopic observations are obtained in a series of at least three 15-minute exposures.
Additional exposures are taken until the squared signal-to-noise ratio per pixel, (S/N)$^2$, reaches the survey-quality threshold for each CCD.
These thresholds are ${\rm (S/N)^2} \geq 22$ at $i$-band magnitude for the red camera
and ${\rm (S/N)^2} \geq 10$ at $g$-band magnitude for the blue camera (Galactic extinction-corrected magnitudes). 
The spectroscopic reduction pipeline for the BOSS spectra is described  
in \cite{bolton2012}.
SDSS-IV uses plates covered by \numprint{1000} fibers that have a field of view of approximately $7 \ {\rm deg^2}$.
The plates are tiled in a manner which allows them to overlap \citep{dawson2016}. \Fig{SkyCoverage} shows the locations of observed plates.
The total area covered by the Data Release 14 of SDSS-IV/eBOSS is \numprint{2044} ${\rm deg^2}$.
\Fig{ProgressPlot} presents the number of spectroscopically confirmed quasars with respect to their observation date.

\section{Construction of the DR14Q catalog}
\label{s:construction}

Unlike the SDSS-III/BOSS quasar catalogs \citep{paris2012,paris2014,paris2017}, the SDSS-IV quasar catalog also contains all the quasars observed as part of SDSS-I/II/III.
This decision is driven by one of the scientific goals of SDSS-IV/eBOSS to use quasars as the tracers of large scale structures at $z \sim 1.5$ \citep[see ][]{dawson2016,blanton2017}: quasars observed as part of the first two iterations of SDSS with a high-quality spectrum, i.e. a spectrum from which one can measure a redshift, were not re-observed as part of SDSS-IV \citep[see ][ for further details]{myers2015}.

\subsection{Definition of the superset}
\label{s:def_superset}

The ultimate goal of the SDSS quasar catalog is to gather all the quasars observed as part of any of the stages of SDSS \citep{york2000,eisenstein2011,blanton2017}.
To do so, we need to create a list of quasar targets as complete as possible that we refer to as the \textit{superset}.
Its definition for the DR14Q catalog depends on the iteration of the SDSS during which a quasar was observed:
\begin{itemize}
	\item SDSS-I/II: we use the list of confirmed quasars in the SDSS-DR7 quasar catalog that contains all spectroscopically confirmed quasars from SDSS-I/II \citep{schneider2010}. A total of \numprint{79487} quasars with no re-observation in SDSS-III/IV are included in the superset for DR14Q.
	\item SDSS-III/IV: we follow the definition of the superset as in \cite{paris2017}. Our input list of quasar targets is composed of all quasar targets as defined by their target selection bits. The full list of programs targeting quasars and associated references is given in \Tab{qts_bit}. This set contains objects targeted as part of the legacy programs but also all the ancillary programs that targeted quasars for specific projects \citep[see e.g.,  ][ for examples of ancillary programs]{dawson2013}. A total of \numprint{819611} quasar targets are identified using target selection bits described in \Tab{qts_bit}.
\end{itemize}
The superset we obtain contains \numprint{899098} objects to be classified.

\begin{table*}
\centering
\begin{tabular}{c l | c l | c l | c l}
\hline
\hline
Bit & Selection & Bit & Selection & Bit & Selection & Bit & Selection \\
\hline
\hline
\multicolumn{8}{l}{BOSS\_TARGET1} \\
\hline
10  & {\tt QSO\_CORE} $^{\rm (1)}$     & 11  & {\tt QSO\_BONUS} $^{\rm (1)}$  & 12  & {\tt QSO\_KNOWN\_MIDZ} $^{\rm (1)}$ & 13  & {\tt QSO\_KNOWN\_LOHIZ} $^{\rm (1)}$ \\
14  & {\tt QSO\_NN} $^{\rm (1)}$  & 15  & {\tt QSO\_UKIDSS} $^{\rm (1)}$ & 16 & {\tt QSO\_LIKE\_COADD} $^{\rm (1)}$ & 17 & {\tt QSO\_LIKE} $^{\rm (1)}$ \\
18 & {\tt QSO\_FIRST\_BOSS} $^{\rm (1)}$ & 19 & {\tt QSO\_KDE} $^{\rm (1)}$ & 40 & {\tt QSO\_CORE\_MAIN} $^{\rm (1)}$ & 41 & {\tt QSO\_BONUS\_MAIN} $^{\rm (1)}$ \\  
42 & {\tt QSO\_CORE\_ED} $^{\rm (1)}$ & 43 & {\tt QSO\_CORE\_LIKE} $^{\rm (1)}$ & 44 & \multicolumn{3}{l}{{\tt QSO\_KNOWN\_SUPPZ} $^{\rm (1)}$} \\
\hline
\multicolumn{8}{l}{ANCILLARY\_TARGET1} \\
\hline
6 & {\tt BLAZGVAR} $^{\rm (2)}$ & 7 & {\tt BLAZR} $^{\rm (2)}$ & 8 & {\tt BLAZXR} $^{\rm (2)}$ & 9 & {\tt BLAZXRSAL} $^{\rm (2)}$ \\
10 & {\tt BLAZXRVAR} $^{\rm (2)}$ & 11 & {\tt XMMBRIGHT} $^{\rm (2)}$ & 12 & {\tt XMMGRIZ} $^{\rm (2)}$ & 13 & {\tt XMMHR} $^{\rm (2)}$ \\
14 & {\tt XMMRED} $^{\rm (2)}$ & 15 & {\tt FBQSBAL} $^{\rm (2)}$ & 16 & {\tt LBQSBAL} $^{\rm (2)}$ & 17 & {\tt ODDBAL} $^{\rm (2)}$ \\
18 & {\tt OTBAL} $^{\rm (2)}$ & 19 & {\tt PREVBAL} $^{\rm (2)}$ & 20 & {\tt VARBAL} $^{\rm (2)}$ & 22 & {\tt QSO\_AAL} $^{\rm (2)}$ \\
23 & {\tt QSO\_AALS} $^{\rm (2)}$ & 24 & {\tt QSO\_IAL} $^{\rm (2)}$ & 25 & {\tt QSO\_RADIO} $^{\rm (2)}$ & 26 & {\tt QSO\_RADIO\_AAL} $^{\rm (2)}$ \\
27 & {\tt QSO\_RADIO\_IAL} $^{\rm (2)}$ & 28 & {\tt QSO\_NOAALS} $^{\rm (2)}$ & 29 & {\tt QSO\_GRI} $^{\rm (2)}$ & 30 & {\tt QSO\_HIZ} $^{\rm (2)}$ \\
31 & {\tt QSO\_RIZ} $^{\rm (2)}$ & 50 & {\tt BLAZGRFLAT} $^{\rm (2)}$ & 51 & {\tt BLAZGRQSO} $^{\rm (2)}$ & 52 & {\tt BLAZGX} $^{\rm (2)}$ \\
53 & {\tt BLAZGXQSO} $^{\rm (2)}$ & 54 & {\tt BLAZGXR} $^{\rm (2)}$ & 55 & {\tt BLAZXR} $^{\rm (2)}$ & 58 & {\tt CXOBRIGHT} $^{\rm (2)}$ \\ 
59 & \multicolumn{7}{l}{{\tt CXORED} $^{\rm (2)}$}\\
\hline
\multicolumn{8}{l}{ANCILLARY\_TARGET2} \\
\hline
0 & {\tt HIZQSO82} $^{\rm (2)}$ & 1 & {\tt HIZQSOIR} $^{\rm (2)}$ & 2 & {\tt KQSO\_BOSS} $^{\rm (2)}$ & 3 & {\tt QSO\_VAR} $^{\rm (2)}$ \\
4 & {\tt QSO\_VAR\_FPG} $^{\rm (2)}$ & 5 & {\tt RADIO\_2LOBE\_QSO} $^{\rm (2)}$ & 7 & {\tt QSO\_SUPPZ} $^{\rm (2)}$ & 8 & {\tt QSO\_VAR\_SDSS} $^{\rm (2)}$ \\
9 & {\tt QSO\_WISE\_SUPP} $^{\rm (3)}$ & 10 & {\tt QSO\_WISE\_FULL\_SKY} $^{\rm (4)}$ & 13 & {\tt DISKEMITTER\_REPEAT} $^{\rm (4)}$ & 14 & {\tt WISE\_BOSS\_QSO} $^{\rm (4)}$ \\
15 & {\tt QSO\_XD\_KDE\_PAIR} $^{\rm (4)}$ & 24 & {\tt TDSS\_PILOT} $^{\rm (4)}$ & 25 & {\tt SPIDERS\_PILOT} $^{\rm (4)}$ & 26 & {\tt TDSS\_SPIDERS\_PILOT} $^{\rm (4)}$ \\
27 & {\tt QSO\_VAR\_LF} $^{\rm (4)}$ & 31 & {\tt QSO\_EBOSS\_W3\_ADM} $^{\rm (4)}$ & 32 & {\tt XMM\_PRIME} $^{\rm (4)}$ & 33 & {\tt XMM\_SECOND} $^{\rm (4)}$ \\
53 & {\tt SEQUELS\_TARGET} $^{\rm (4)}$ & 54 & {\tt RM\_TILE1} $^{\rm (4)}$ & 55 & {\tt RM\_TILE2} $^{\rm (4)}$ & 56 & {\tt QSO\_DEEP} $^{\rm (4)}$ \\
\hline
\multicolumn{8}{l}{EBOSS\_TARGET0} \\
\hline
10 & {\tt QSO\_EBOSS\_CORE} $^{\rm (5)}$ & 11 & {\tt QSO\_PTF} $^{\rm (5)}$ & 12 & {\tt QSO\_REOBS} $^{\rm (5)}$ & 13 & {\tt QSO\_EBOSS\_KDE} $^{\rm (5)}$ \\
14 & {\tt QSO\_EBOSS\_FIRST} $^{\rm (5)}$ & 15 & {\tt QSO\_BAD\_BOSS} $^{\rm (5)}$ & 16 & {\tt QSO\_BOSS\_TARGET} $^{\rm (5)}$ & 17 & {\tt QSO\_SDSS\_TARGET} $^{\rm (5)}$ \\
18 & {\tt QSO\_KNOWN} $^{\rm (5)}$ & 20 & {\tt SPIDERS\_RASS\_AGN} $^{\rm (6)}$ & 22 & {\tt SPIDERS\_ERASS\_AGN} $^{\rm (6)}$ & 30 & {\tt TDSS\_A} $^{\rm (7)}$ \\
31 & {\tt TDSS\_FES\_DE} $^{\rm (8)}$ & 33 & {\tt TDSS\_FES\_NQHISN} $^{\rm (8)}$ & 34 & {\tt TDSS\_FES\_MGII} $^{\rm (8)}$ & 35 & {\tt TDSS\_FES\_VARBAL} $^{\rm (8)}$ \\
40 & \multicolumn{7}{l}{\tt SEQUELS\_PTF\_VARIABLE}\\
\hline
\multicolumn{8}{l}{EBOSS\_TARGET1} \\
\hline
9 & {\tt QSO1\_VAR\_S82} $^{\rm (9)}$ & 10 & {\tt QSO1\_EBOSS\_CORE} $^{\rm (5)}$ & 11 & {\tt QSO1\_PTF} $^{\rm (5)}$ & 12 & {\tt QSO1\_REOBS} $^{\rm (5)}$ \\
13 & {\tt QSO1\_EBOSS\_KDE} $^{\rm (5)}$ & 14 & {\tt QSO1\_EBOSS\_FIRST} $^{\rm (5)}$ & 15 & {\tt QSO1\_BAD\_BOSS} $^{\rm (5)}$ & 16 & {\tt QSO\_BOSS\_TARGET} $^{\rm (5)}$ \\
17 & {\tt QSO\_SDSS\_TARGET} $^{\rm (5)}$ & 18 & {\tt QSO\_KNOWN} $^{\rm (5)}$ & 30 & {\tt TDSS\_TARGET} $^{\rm (7, 8, 10)}$ & 31 & {\tt SPIDERS\_TARGET} $^{\rm (6, 10)}$ \\
\hline
\multicolumn{8}{l}{EBOSS\_TARGET2} \\
\hline
0 & {\tt SPIDERS\_RASS\_AGN} $^{\rm (6)}$ & 2 & {\tt SPIDERS\_ERASS\_ANG} $^{\rm (6)}$ & 4 & {\tt SPIDERS\_XMMSL\_AGN} $^{\rm (6)}$ & 20 & {\tt TDSS\_A} $^{\rm (7)}$ \\
21 & {\tt TDSS\_FES\_DE} $^{\rm (8)}$ & 23 & {\tt TDSS\_FES\_NQHISN} $^{\rm (8)}$ & 24 & {\tt TDSS\_FES\_MGII} $^{\rm (8)}$ & 25 & {\tt TDSS\_FES\_VARBAL} $^{\rm (8)}$ \\
26 & {\tt TDSS\_B} $^{\rm (7)}$ & 27 & {\tt TDSS\_FES\_HYPQSO} $^{\rm (8)}$ & 31 & {\tt TDSS\_CP} $^{\rm (8)}$ & 32 & {\tt S82X\_TILE1} $^{\rm (10)}$ \\
33 & {\tt S82X\_TILE2} $^{\rm (10)}$ & 34 & {\tt S82X\_TILE3} $^{\rm (10)}$ & 50 & {\tt S82X\_BRIGHT\_TARGET} $^{\rm (10)}$ & 51 & {\tt S82X\_XMM\_TARGET} $^{\rm (10)}$ \\
52 & \multicolumn{3}{l|}{{\tt S82X\_WISE\_TARGET} $^{\rm (10)}$} & 53 & \multicolumn{3}{l}{{\tt S82X\_SACLAY\_VAR\_TARGET} $^{\rm (10)}$} \\  
54 & \multicolumn{3}{l|}{{\tt S82X\_SACLAY\_BDT\_TARGET} $^{\rm (10)}$} & 55 & \multicolumn{3}{l}{{\tt S82X\_SACLAY\_HIZ\_TARGET} $^{\rm (10)}$} \\
56 & \multicolumn{3}{l|}{{\tt S82X\_RICHARDS15\_PHOTOQSO\_TARGET} $^{\rm (10)}$} & 57 & \multicolumn{3}{l}{{\tt S82X\_PETERS15\_COLORVAR\_TARGET} $^{\rm (10)}$} \\
58 & \multicolumn{3}{l|}{{\tt S82X\_LSSTZ4\_TARGET} $^{\rm (10)}$} & 59 & \multicolumn{3}{l}{{\tt S82X\_UNWISE\_TARGET} $^{\rm (10)}$} \\ 
60 & \multicolumn{3}{l|}{{\tt S82X\_GTRADMZ4\_TARGET} $^{\rm (10)}$} & 61 & \multicolumn{3}{l}{{\tt S82X\_CLAGN1\_TARGET} $^{\rm (10)}$} \\
62 & \multicolumn{7}{l}{{\tt S82X\_CLAGN2\_TARGET} $^{\rm (10)}$} \\
\hline
\multicolumn{8}{l}{$^{\rm (1)}$ \cite{ross2012} -- $^{\rm (2)}$ \cite{dawson2013} -- $^{\rm (3)}$ \cite{DR10} -- $^{\rm (4)}$ \cite{DR12} -- $^{\rm (5)}$ \cite{myers2015}} \\
\multicolumn{8}{l}{$^{\rm (6)}$ \cite{dwelly2017} -- $^{\rm (7)}$ \cite{morganson2015} -- $^{\rm (8)}$ \cite{macleod2017} -- $^{\rm (9)}$ \cite{palanque2016}} \\
\multicolumn{8}{l}{$^{\rm (10)}$ \cite{DR14}} \\
\hline
\end{tabular}
\caption{List of programs from SDSS-III/BOSS \citep{eisenstein2011,dawson2013} and SDSS-IV \citep{dawson2016,blanton2017} targeting quasars and taken as inputs to build up the SDSS-DR14 quasar catalog.
}
\label{t:qts_bit}
\end{table*}

\subsection{Automated classification}
\label{s:def_autoclass}

Given the increase of the number of quasar targets in SDSS-IV, the systematic visual inspection we performed in SDSS-III/BOSS \citep[e.g. ][]{paris2012}  is no longer feasible. 
Since the output of the SDSS pipeline \citep{bolton2012} cannot be fully efficient to classify quasar targets, we adopt an alternate strategy: starting from the output of the SDSS pipeline, we identify SDSS-IV quasar targets for which it is likely the identification and redshifts are inaccurate. 
This set of objects is visually inspected following the procedure described in \Sec{vi_process}.

The spectra of quasar candidates are reduced by the SDSS pipeline\footnote{The software used to reduce SDSS
data is called idlspec2d. Its DR14 version is v5\_10\_0.}, which provides a classification ({\tt QSO}, {\tt STAR} or {\tt GALAXY}) and a redshift.
This task is accomplished using a library of stellar templates and a principal component analysis (PCA) decomposition of galaxy and quasar spectra are fitted to each spectrum.
Each class of templates is fitted in a given range of redshift: 
galaxies from $z = -0.01$ to $1.00$, quasars from $z = 0.0033$ to $7.00$, and 
stars from $z = -0.004$ to $0.004$ ($\pm$\numprint{1200}$\ {\rm km \ s^{-1}}$).
For each spectrum, the fits are ordered by increasing reduced $\chi ^2$; the overall best fit is the fit with the lowest reduced $\chi ^2$.

We start with the first five identifications, i.e. identifications corresponding to the five lowest reduced $\chi ^2$, redshifts and {\tt ZWARNING}. The latter is a quality flag. Whenever it is set to 0, its classification and redshift are considered reliable. We then apply the following algorithm:
\begin{enumerate}
	\item If the first SDSS pipeline identification is {\tt STAR}, then the resulting classification is {\tt STAR};
	\item If the first SDSS pipeline identification is {\tt GALAXY} with $z_{pipeline} < 1$, then the resulting classification is {\tt GALAXY};
	\item If the first SDSS pipeline identification is {\tt GALAXY} with $z_{pipeline} \leq 1$ \textit{and} at least two other SDSS pipeline identifications are {\tt GALAXY}, then the resulting classification is {\tt GALAXY};
	\item If the first SDSS pipeline identification is {\tt QSO} with {\tt ZWARNING}~=~0, then the resulting classification is {\tt QSO}, \textit{except} if at least two other SDSS pipeline identifications are {\tt STAR}. In such a case, the resulting identification is {\tt STAR};
	\item If the first pipeline identification is {\tt QSO} with {\tt ZWARNING}$>0$ and at least two alternate SDSS pipeline identifications are {\tt STAR}, then the resulting identification is {\tt STAR}.
\end{enumerate}
At this stage, the redshift measurement we consider for automatically classified objects is the redshift estimate of the overall best fit of the SDSS pipeline, except if the automated identification is {\tt STAR}. 
In that case, we set the redshift to 0.
If an object does not pass any of these conditions, the resulting classification is {\tt UNKNOWN} and it is added to the list of objects that require visual inspection (see \Sec{vi_process}).
\\

In order to achieve the expected precision on the $d_A \left( z \right)$ and $H \left( z \right)$ measurements, it is required (i) to have less than 1\% of actual quasars lost in the classification process and (ii) to have less than 1\% of contaminants in the quasar catalog.
We tested this algorithm against the result of the full visual inspection of the ``SEQUELS'' pilot survey of SDSS-IV/eBOSS that contains a total of \numprint{36489} objects \citep[see ][ for details]{myers2015,paris2017} to ensure that these requirements are fulfilled.

Out of these \numprint{36489} objects, \numprint{2393} (6.6\% of the whole sample) cannot be classified by the automated procedure, \numprint{18799} are classified as {\tt QSO}, \numprint{10001} as {\tt STAR}, and \numprint{5288} as {\tt GALAXY}.
For objects identified as {\tt QSO} by the algorithm, \numprint{98} are wrongly classified. This represents a contamination of the quasar sample of 0.5\%.
A total of \numprint{158} actual quasars, i.e. identified in the course of the full visual inspection, are lost which represents 0.8\% of the whole quasar sample. 
The latter number includes 12 objects identified as {\tt QSO\_Z?} by visual inspection because their identification is not ambiguous.
Detailed results for the comparison with the fully visually inspected sample are provided in \Tab{efficiency_autoclass}.

The performance of this algorithm depends on the SDSS pipeline version and the overall data quality.
To ensure that this performance does not change significantly, we fully visually inspect randomly picked plates regularly and test the quality of the output.

\begin{table*}
\centering
\begin{tabular}{l r r r r}
\hline
\hline
                         & \multicolumn{4}{c}{Automated classification} \\
                         & {\tt UNKNOWN}       & {\tt QSO}         & {\tt STAR}      & {\tt GALAXY}    \\ 
Visual inspection        &                     &                   &                 &                 \\
\hline
\hline
{\tt QSO}                & \numprint{1418}     & \numprint{17687}  & \numprint{27}   & \numprint{107}  \\
{\tt QSO\_BAL}           & \numprint{144}      & \numprint{946}    & \numprint{4}    & \numprint{8}    \\
{\tt QSO\_?}             & \numprint{79}       & \numprint{27}     & \numprint{6}    & \numprint{26}   \\
{\tt QSO\_Z?}            & \numprint{120}      & \numprint{41}     & \numprint{5}    & \numprint{7}    \\
{\tt Star}               & \numprint{213}      & \numprint{11}     & \numprint{9804} & \numprint{189}  \\
{\tt Star\_?}            & \numprint{61}       & \numprint{5}      & \numprint{54}   & \numprint{29}   \\
{\tt Galaxy}             & \numprint{74}       & \numprint{37}     & \numprint{23}   & \numprint{4623} \\
{\tt ?}                  & \numprint{115}      & \numprint{15}     & \numprint{34}   & \numprint{95}   \\
{\tt Bad}                & \numprint{169}      & \numprint{30}     & \numprint{44}   & \numprint{204}  \\
\hline
\end{tabular}
\caption{Comparison between the automated classification scheme based on the output of the SDSS pipeline \citep{bolton2012} and visual inspection of our training set.
}
\label{t:efficiency_autoclass}
\end{table*}

\subsection{Visual inspection process}
\label{s:vi_process}

Depending on the iteration of SDSS, different visual inspection strategies have been applied.

	\subsubsection{Systematic visual inspection of SDSS-III quasar candidates}
	
As described in \cite{paris2012}, we visually inspected all quasar targets observed in SDSS-III/BOSS \citep{eisenstein2011,dawson2013}. The idea was to construct a
quasar catalog as complete and pure as possible (this was also the approach adopted by SDSS-I/II catalogs). Several checks during the SDSS-III/BOSS survey have shown that completeness (within the given target selection) and purity are larger
than 99.5\%.

After their observation, all the spectra are automatically classified by the SDSS pipeline \citep{bolton2012}.
Spectra are divided into four categories based on their initial classification by the SDSS pipeline: low-redshift quasars (i.e. $z < 2$), high-redshift quasars (i.e. $z \geq 2$), stars and others.
We perform the visual inspection plate by plate through a dedicated website: all spectra for a given category can be validated at once if their identification and redshift are correct.
If an object requires further inspection or a change in its redshift, we have the option to go to a detailed page on which not only the identification can be changed but
also BALs and DLAs can be flagged and the redshift can be adjusted. When possible
the peak of the \ion{Mg}{ii} emission line was used as an estimator of the redshift
\citep[see ][]{paris2012}, otherwise the peak of \ion{C}{iv} was taken as the indicator in case the redshift given by the pipeline was obviously in error.

	\subsubsection{Residual visual inspection of SDSS-IV quasar candidates}

For SDSS-IV/eBOSS, we visually inspect only the objects the automated procedure considers
ill-identified. Most of the corresponding spectra are, unsurprisingly, of low S/N. A number of ill-identified sources have good S/N but show strong absorption lines which confuse the pipeline. 
These objects can be strong BALs but also spectra with a strong DLA at the emission 
redshift \citep{finley2013,fathivavsari2017}. A few objects have very unusual continua. 
The visual inspection itself proceeds as for the SDSS-III/BOSS survey. However, we no longer visually flag BALs and we change redshifts only in case of catastrophic failures of the SDSS pipeline.

\subsection{Classification result}
\label{s:res_class}

Starting from the \numprint{899098} unique objects included in the DR14Q superset, we run the automated procedure described in \Sec{def_autoclass}.
A total of \numprint{42729} quasar candidates are classified as {\tt UNKNOWN} by the algorithm.
Identification from the full visual inspection of SDSS-I/II/III quasar targets was already available for \numprint{625432} objects, including \numprint{32621} identified as {\tt UNKNOWN} by the automated procedure.
The remaining \numprint{10108} quasar candidates with no previous identification from SDSS-I/II/III have been visually inspected.
After merging all the already existing identifications, \numprint{635540} objects are identified through visual inspection and \numprint{263558} are identified by the automated procedure.

A total of \numprint{526356} quasars are identified, \numprint{387223} from visual inspection and \numprint{139133} from the automated classification.
Results are summarized in \Tab{res_class}.

\begin{table*}
\centering
\begin{tabular}{l r}
\hline
\hline
Number of quasar targets in SDSS-III/IV             & \numprint{819611} \\
Number of quasars from SDSS-I/II only               & \numprint{79487}   \\
\textbf{Total number of objects in DR14Q superset}  & \textbf{\numprint{899098}} \\
\hline
Number of targets with automated classification     & \numprint{263558} \\
Number of targets with visual inspection classification & \numprint{635540} \\
\hline
Number of quasars from automated classification     & \numprint{139133} \\
Number of quasars from visual inspection classification & \numprint{387223} \\
\textbf{Total number of DR14Q quasars}              & \textbf{\numprint{526356}} \\
\hline
\end{tabular}
\caption{
Result of our classification process, split between automated and visual inspection classifications.
}
\label{t:res_class}
\end{table*}    

\section{Redshift estimate}
\label{s:redshift}

Despite the presence of large and prominent emission lines, it is frequently difficult to estimate accurate redshifts for quasars.
Indeed, the existence of quasar outflows create systematic shifts in the location of broad emission lines leading to not fully controlled errors in the measurement of redshifts \citep[e.g. ][]{shen2016}.
Accuracy in this measurement is crucial to achieve the scientific goals of SDSS-IV/eBOSS.
As stated in the Sec. 5.2 of \cite{dawson2016}, we mitigate this problem by using two different types of redshift estimates: one based on the result of a principal component analysis and another one based on the location of the maximum of the peak of the \ion{Mg}{ii} emission line.

	\subsection{Automated redshift estimates}
	\label{s:zauto}	
    
Various studies have shown that the \ion{Mg}{ii} emission line is the quasar broad emission line that is the least affected by systematic shifts \citep[e.g. ][]{hewett2010,shen2016}.
In the BOSS spectral range, this feature is available in the redshift range 0.3--2.5, which covers most of our sample.
\\

To measure the \ion{Mg}{ii} redshift ({\tt Z\_MGII}), we first perform a principal component analysis (PCA) on a sample of \numprint{ 8986} SDSS-DR7 quasars \citep{schneider2010} using input redshifts from \cite{hewett2010}. The detailed selection of this sample is explained in Sec. 4 of \cite{paris2012}.
With the resulting set of eigenspectra, we fit a linear combination of five principal components and measure the location of the maximum of the \ion{Mg}{ii} emission line.
This first step produces a new redshift measurement that can be used to re-calibrate our reference sample.
We then perform another PCA with {\tt Z\_MGII} and derive a new set of principal components.
In this second step it is not necessary to have {\tt Z\_MGII} but this step is mandatory to derive PCA redshifts calibrated to use the \ion{Mg}{ii} emission as a reference. 

Finally, to measure {\tt Z\_PCA}, we fit a linear combination of four eigenspectra to all DR14Q spectra.
The redshift estimate is an additional free parameter in the fit.
During the fitting process, there is an iterative removal of absorption lines in order to limit their impact on redshift measurements; Details are given in \cite{paris2012}.

	\subsection{Comparison of redshift estimates provided in DR14Q}

\begin{figure*}[htbp]
	\centering{\includegraphics[width=\linewidth]{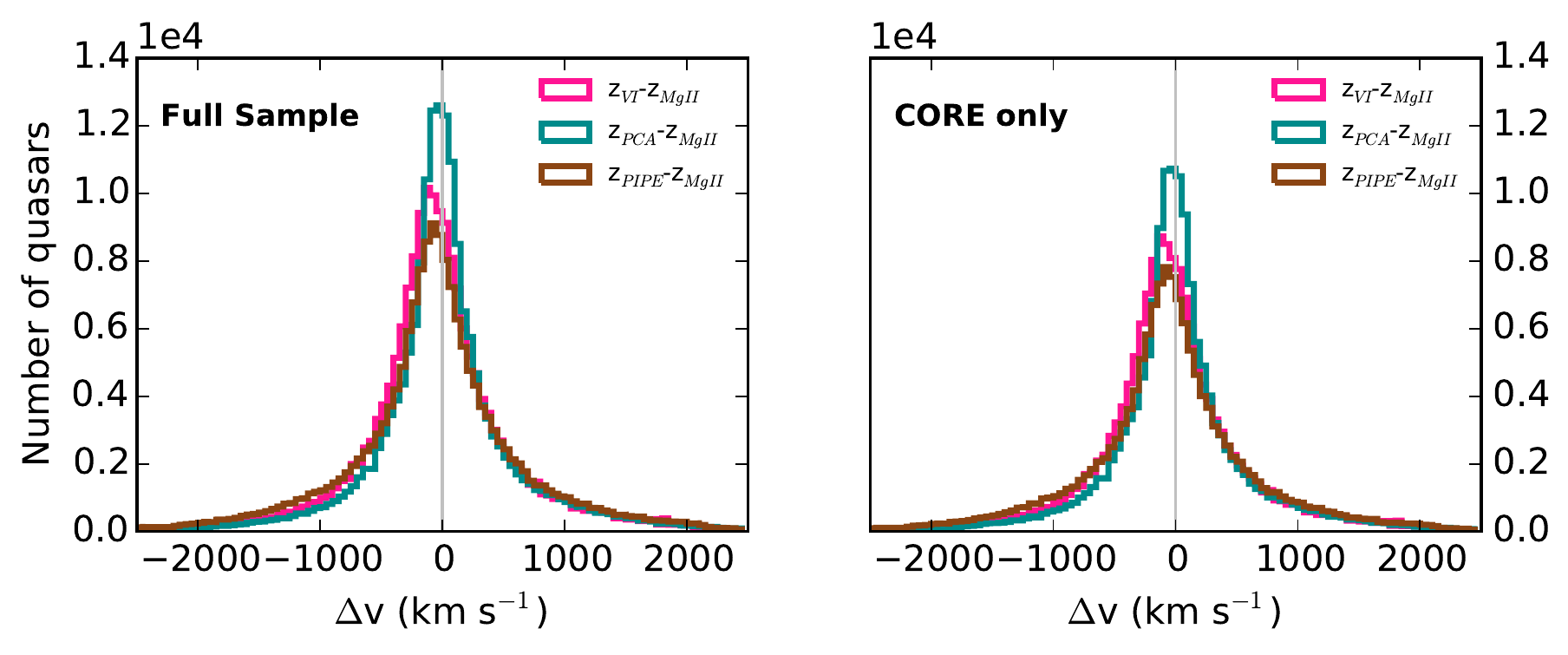}}
\caption{ 
\textit{Left Panel: }
Distribution of velocity differences between {\tt Z\_VI} (magenta histogram), {\tt Z\_PIPE} (brown histogram), {\tt Z\_PCA} (dark cyan histogram) and {\tt Z\_MGII}.
These histograms are computed using \numprint{178981} quasars that have all the four redshift estimates available in DR14Q.
\textit{Right Panel: }
Same as left panel restricted to the \numprint{151701} {\tt CORE} quasars for which we have the four redshift estimates available in DR14Q.
In both panels, the histograms are computed in bins of $\Delta v = 50 \ {\rm \kms}$ and the vertical grey line marks $\Delta v = 0 \ {\rm km \ s^{-1}}$.
}
\label{fig:redshift_error}
\end{figure*}

In the present catalog, we release four redshift estimates: {\tt Z\_PIPE}, {\tt Z\_VI}, {\tt Z\_PCA}, and {\tt Z\_MGII}.
As explained in the previous section, the \ion{Mg}{ii} emission line is the least affected broad emission line in quasar spectra. In addition, this emission line is available for most of our sample. We use it as the reference redshift to test the accuracy of our three other redshift estimates.
For this test, we select all the DR14Q quasars for which we have the four redshift estimates. Among these \numprint{178981} objects, we also select \numprint{151701} {\tt CORE} quasars only to test the behavior of our estimates on this sample for which redshift accuracy is crucial.
\Fig{redshift_error} displays the distribution of the velocity differences between {\tt Z\_VI}, {\tt Z\_PIPE}, {\tt Z\_PCA} and {\tt Z\_MGII} for the full sample having the four redshift estimates available (left panel) and {\tt CORE} quasars only (right panel). 
\Tab{redshift_accuracy} gives the systematic shift for each of the distributions and the dispersion of these quantities, expressed in ${\rm km \ s^{-1}}$, for both samples.
\\

As explained in \Sec{vi_process}, the visual inspection redshift {\tt Z\_VI} is set to be at the location of the maximum of the \ion{Mg}{ii} emission line when this line is available. With this strategy, the systematic shift with respect to the \ion{Mg}{ii} emission line is limited by the accuracy of the visual inspection. Although it is a time-consuming approach, this redshift estimate produces an extremely low number of redshift failures (less than 0.5\%), leading to a low dispersion around this systematic shift.

The SDSS pipeline redshift estimate, {\tt Z\_PIPE}, is the result of a principal component analysis performed on a sample of visually-inspected quasars. Hence, {\tt Z\_PIPE} is expected to have a similar systematic shift as {\tt Z\_VI}. On the other hand, {\tt Z\_PIPE} is subject to more redshift failures due to peculiar objects or low S/N spectra and thus the larger dispersion of the velocity difference distribution seen in \Fig{redshift_error} and \Tab{redshift_accuracy}.

{Z\_PCA} is also the result of a principal component analysis but, unlike {\tt Z\_PIPE}, the reference sample has been carefully chosen to have an automated redshift corresponding to the location of the maximum of the \ion{Mg}{ii} emission line. Therefore, a systematic shift smaller than $10 \ {\rm km \ s^{-1}}$ was expected when compared to {\tt Z\_MGII}.
In addition, {\tt Z\_PCA} takes into account the possible presence of absorption lines, even broad ones, and it is trained to ignore them. {\tt Z\_PCA} is thus less sensitive to peculiarities in quasar spectra, which explains the reduced dispersion of redshift errors when compared to {\tt Z\_PIPE}.
\\

A similar analysis performed on a sample of \numprint{151701} {\tt CORE} quasars for which we have the four redshift estimates leads to similar results for redshift estimates.
These exercises demonstrate that there is no additional and significant systematics for the redshift estimate of the {\tt CORE} quasar sample.

\begin{table}
\centering
\begin{tabular}{l c c}
\hline
\hline
                             & Median $\Delta v$   &  $\sigma _{\Delta v}$ \\
                             & ($\kms$)            & ($\kms$)              \\
\hline
\hline
\multicolumn{3}{l}{Full sample (\numprint{178981} quasars)} \\
\hline
${\rm z_{VI} - z_{MgII}}$    & -59.7    & 585.9      \\
${\rm z_{PIPE} - z_{MgII}}$  & -54.8    & 697.2      \\
${\rm z_{PCA} - z_{MgII}}$   & -8.8     & 586.9      \\
\hline
\hline
\multicolumn{3}{l}{{\tt CORE} sample (\numprint{151701} quasars)} \\
\hline
${\rm z_{VI} - z_{MgII}}$    & -61.7    & 581.0      \\
${\rm z_{PIPE} - z_{MgII}}$  & -57.9    & 692.1      \\
${\rm z_{PCA} - z_{MgII}}$   & -9.3     & 577.4      \\
\hline
\end{tabular}
\caption{Median shift and dispersion of velocity differences between {\tt Z\_VI}, {\tt Z\_PIPE}, {\tt Z\_PCA} and {\tt Z\_MGII} for the full DR14Q sample and the {\tt CORE} sample only.
These numbers are derived from a subsample of \numprint{178981} quasars from DR14Q for which we have the four redshift estimates available. 
We also restrict this analysis to \numprint{151701} {\tt CORE} quasars for which we also have the four different redshift estimates. See \Fig{redshift_error} for the full distributions.
}
\label{t:redshift_accuracy}
\end{table}

\section{Broad Absorption Line quasars}
\label{s:bal}

In SDSS-III/BOSS, we performed a full visual inspection of all quasar targets.
During this process, we visually flagged spectra displaying broad absorption lines (BAL). 
With this catalog, it is no longer possible to visually inspect all BALs and we now rely on a fully automated detection of BALs.
\\

As for the previous SDSS quasar catalogs, we automatically search for BAL features and report metrics of common use in the community: the BALnicity Index \citep[BI; ][]{weymann1991} of the \ion{C}{iv} absorption troughs.
We restrict the automatic search to quasars with $z \geq 1.57$ in order to have the full spectral 
coverage of \ion{C}{iv} absorption troughs.
The BALnicity index (Col. \#32) is computed bluewards of the \ion{C}{iv} emission line and is defined as:
\begin{equation}
 {\rm BI} ~=~ - \int ^{3000} _{25000} \left[ 1 - \frac{f \left( v \right)}{0.9} \right] C \left( v \right) {\rm d}v ,
 \label{eq:BI_def}
\end{equation}
where $f \left( v \right)$ is the normalized flux density as a function of velocity displacement from the emission-line center. 
The quasar continuum is estimated using the linear combination of four principal components as described in \Sec{zauto}.
$C \left( v \right)$ is initially set to 0 and can take
only two values, 1 or 0. It is set to 1 whenever the quantity $ 1 - f \left( v \right)/0.9$ is continuously positive over an interval of at least \numprint{2000}
$\kms$. It is reset to 0 whenever this quantity becomes negative.
\ion{C}{iv} absorption troughs wider than \numprint{2000} $\kms$ are detected in the spectra of \numprint{21877} quasars. 

\begin{figure}[htbp]
	\centering{\includegraphics[width=\linewidth]{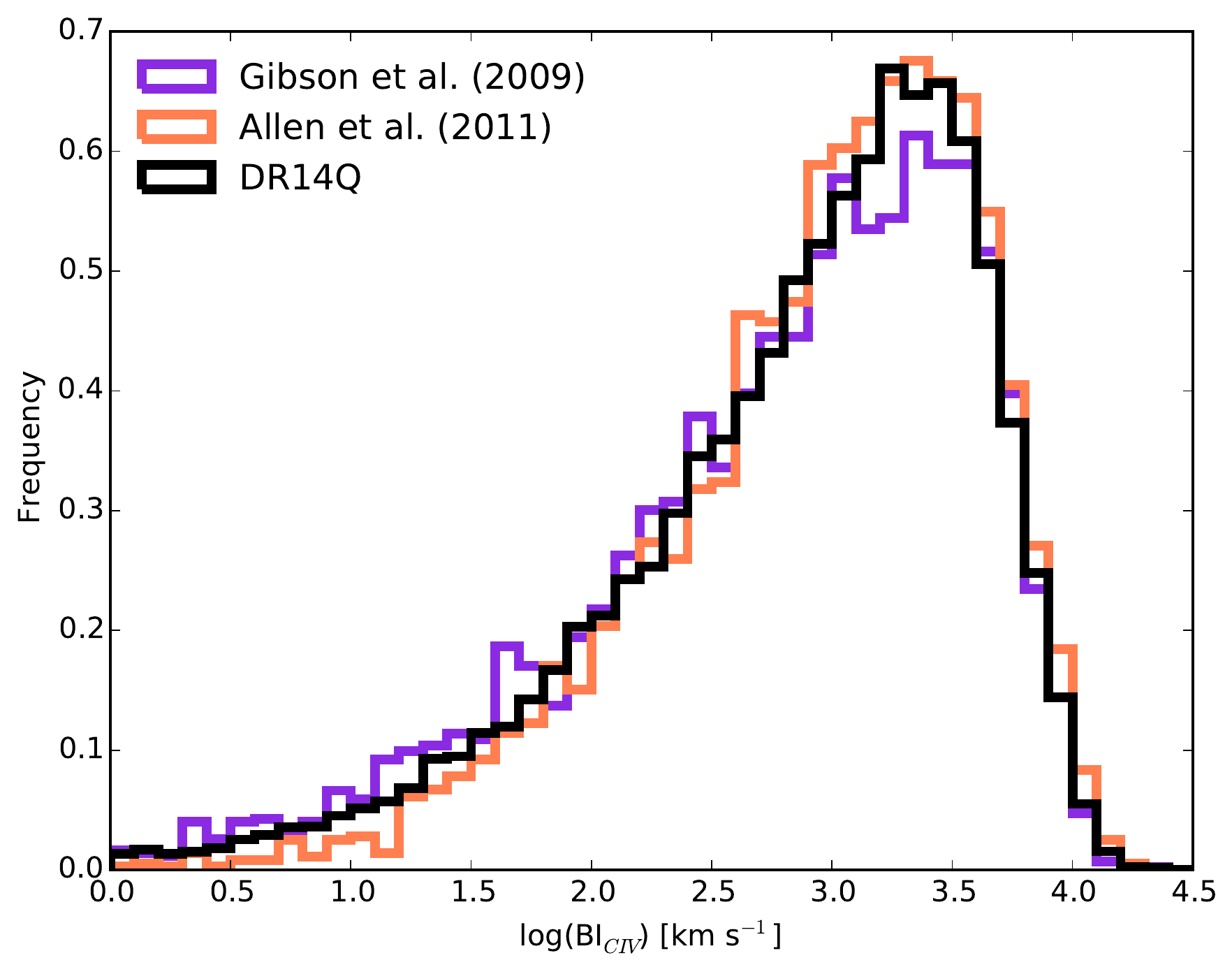}}
\caption{ 
Distribution of the logarithm of the \ion{C}{iv} BALnicity index for BAL quasars in DR14Q (black histogram), \cite[ violet histogram]{gibson09} and \cite[ orange histogram]{allen2011}. All the histograms are normalized to have their sum equal to one. Histograms have bin widths of $\log {\rm BI_{CIV}} = 0.1$.
}
\label{fig:distributionBI}
\end{figure}

The distribution of BI for \ion{C}{iv} troughs from DR14Q is presented in \Fig{distributionBI} (black histogram) and is compared to previous works by \citeauthor{gibson09} \citeyearpar[purple histogram]{gibson09} performed on DR5Q \citep{schneider2007} and 
by \citeauthor{allen2011} \citeyearpar[orange histogram]{allen2011} who searched for BAL quasars in quasar spectra released as part of SDSS-DR6 \citep{DR6}.
The three distributions are normalized to have their sum equal to one. 
The overall shapes of the three distributions are similar. 
The BI distribution from \cite{gibson09} exhibits a slight excess of low-BI values ($\log {\rm BI_{CIV}} < 2$) compared to \cite{allen2011} 
and this work.  The most likely explanation is the difference in the quasar emission modeling. 
\cite{allen2011} used a non-negative matrix factorization (NMF) to estimate the unabsorbed
flux, which produces a quasar emission line shape akin to the one we obtain with PCA. \cite{gibson09} modeled their quasar continuum with a reddened power-law and strong emission lines with Voigt profiles. 
Power-law like continua tend to underestimate the actual quasar emission and hence, the resulting BI values tend to be lower than the one computed when the quasar emission is modeled with NMF or PCA methods.

\section{Summary of the sample}
\label{s:sample}

The DR14Q catalog contains \numprint{526356} unique quasars, of which \numprint{144046} are new discoveries since the previous release.
This dataset represents an increase of about 40\% in the number of SDSS quasars since the beginning of SDSS-IV.
Spectroscopic observations of quasars were performed over \numprint{9376} deg$^2$ for SDSS-I/II/III. New SDSS-IV spectroscopic data are available over \numprint{2044} deg$^2$.
The average surface density of $ 0.9 < z < 2.2$ quasars prior to the beginning of SDSS-IV is 13.27 ${\rm deg ^{-2}}$, and reaches 80.24 ${\rm deg ^{-2}}$ in regions for which SDSS-IV spectroscopy is available.
The overall quasar surface density in regions with SDSS-IV spectroscopy is 125.03 ${\rm deg^{-2}}$, which corresponds to an increase by a factor of 2.4 times compared to the previous quasar catalog release.

\begin{figure*}[htbp]
	\centering{\includegraphics[width=\linewidth]{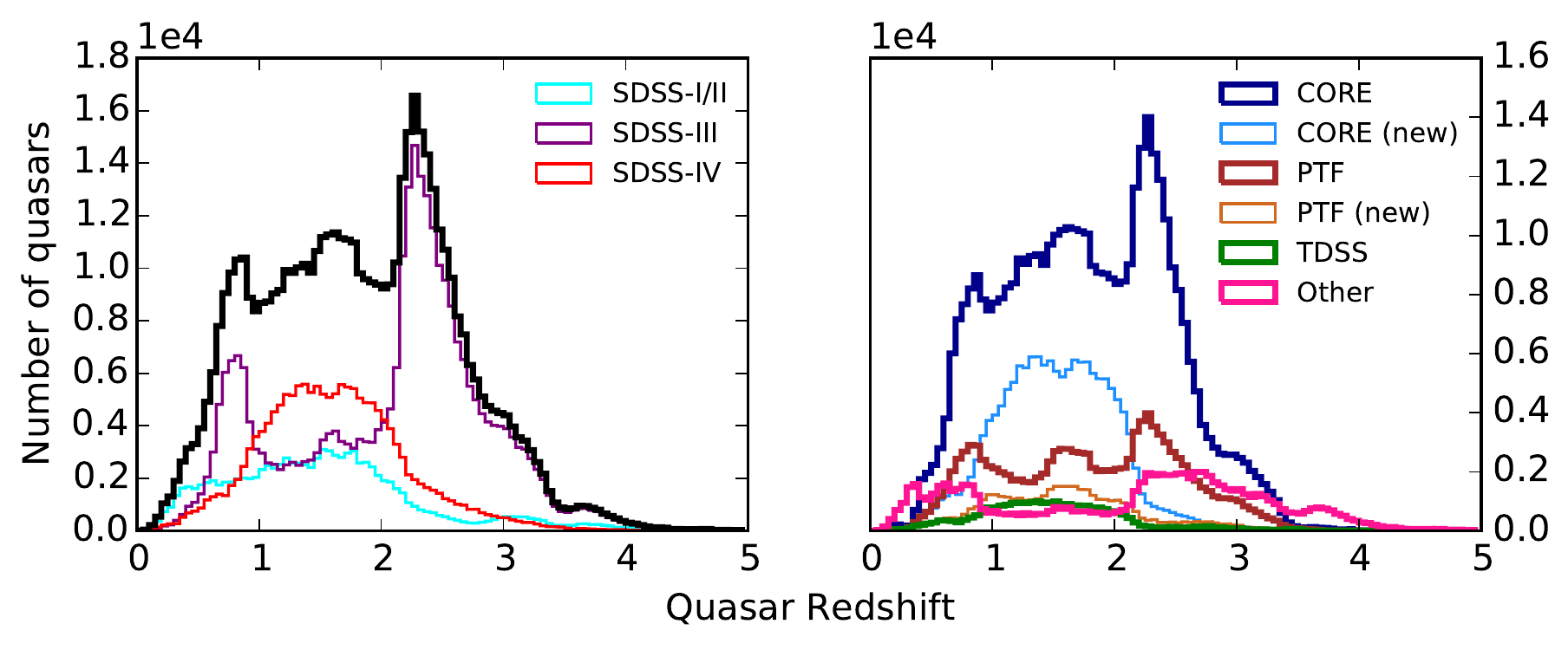}}
\caption{ 
\textit{Left Panel: }
Redshift distribution of DR14Q quasars (black thick histogram) in the range $0 \leq z \leq 5$. The redshift distribution of quasars observed as part of SDSS-I/II is shown with a cyan histogram. 
The redshift distributions of quasars observed as part of SDSS-III (purple histogram) and SDSS-IV (red histogram) are also displayed.
\textit{Right Panel: }
Redshift distributions of all {\tt CORE} quasars that are part of DR14Q (dark blue histogram), {\tt CORE} quasars observed as part of SDSS-IV/eBOSS (light blue histogram), all {\tt PTF} quasars (brown histogram), {\tt PTF} quasars observed as part of SDSS-IV/eBOSS (orange histogram), SDSS-IV/TDSS quasars (green histogram) and quasars observed as part of ancillary programs (pink histogram).
Some quasars can be selected by several target selection algorithms, hence the cumulative number of quasars in a single redshift bin can exceed the total number in that bin.
The bin size for both panels is $\Delta z = 0.05$.
}
\label{fig:zdistri}
\end{figure*}

The redshift distribution of the full sample is shown in \Fig{zdistri} (left panel; black histogram). 
Redshift distributions of quasars observed by each phase of SDSS are also presented in the left panel of \Fig{zdistri}: SDSS-I/II in cyan, SDSS-III in purple and SDSS-IV in red.
SDSS-I/II has observed quasars in the redshift range 0--5.4 with an almost flat distribution up to $z \sim 2.5$, and then a steep decrease.
SDSS-III has focused on $z \geq 2.15$ quasars in order to access the \lya\ forest. The two peaks at $z \sim 0.8$ and $z \sim 1.6$ are due to known degeneracies in the associated quasar target selection \cite[see ][ for more details]{ross2012}.
SDSS-IV/eBOSS  mostly aims to fill in the gap between $z \sim 0.8$ and $z \sim 2$.
It should be noted that some quasars have been observed multiple times throughout the 16 years of the survey, thus the cumulative number in each redshift bin is larger than the number of objects in each redshift bin for the full sample.
The right panel of \Fig{zdistri}
shows the redshift distributions for each of the sub-programs using different
target selection criteria \citep{myers2015}.
The thick blue histogram indicates the redshift distribution of the {\tt CORE} sample taking into account previous spectroscopic observations from SDSS-I/II/III.
The light blue histogram is the redshift distribution of new SDSS-IV {\tt CORE} quasars, i.e. those that have been observed later than July 2014.
The thick brown histogram displays the redshift distribution of all the variability-selected quasars, i.e. including quasars that were spectroscopically confirmed in SDSS-I/II/III.
The orange histogram represents the redshift distribution of newly confirmed variability-selected quasars by SDSS-IV.
The green histogram represents the redshift distribution of quasars that were targeted for recent spectra as TDSS variables \citep{morganson2015,macleod2017}.
Further discussion of the redshift distribution of TDSS-selected quasars can be found in \cite{ruan2016}
All the quasars selected by other programs, such as ancillary programs in SDSS-III or special plates, have their redshift distribution indicated in pink.

\begin{figure*}[htbp]
	\centering{\includegraphics[width=\linewidth]{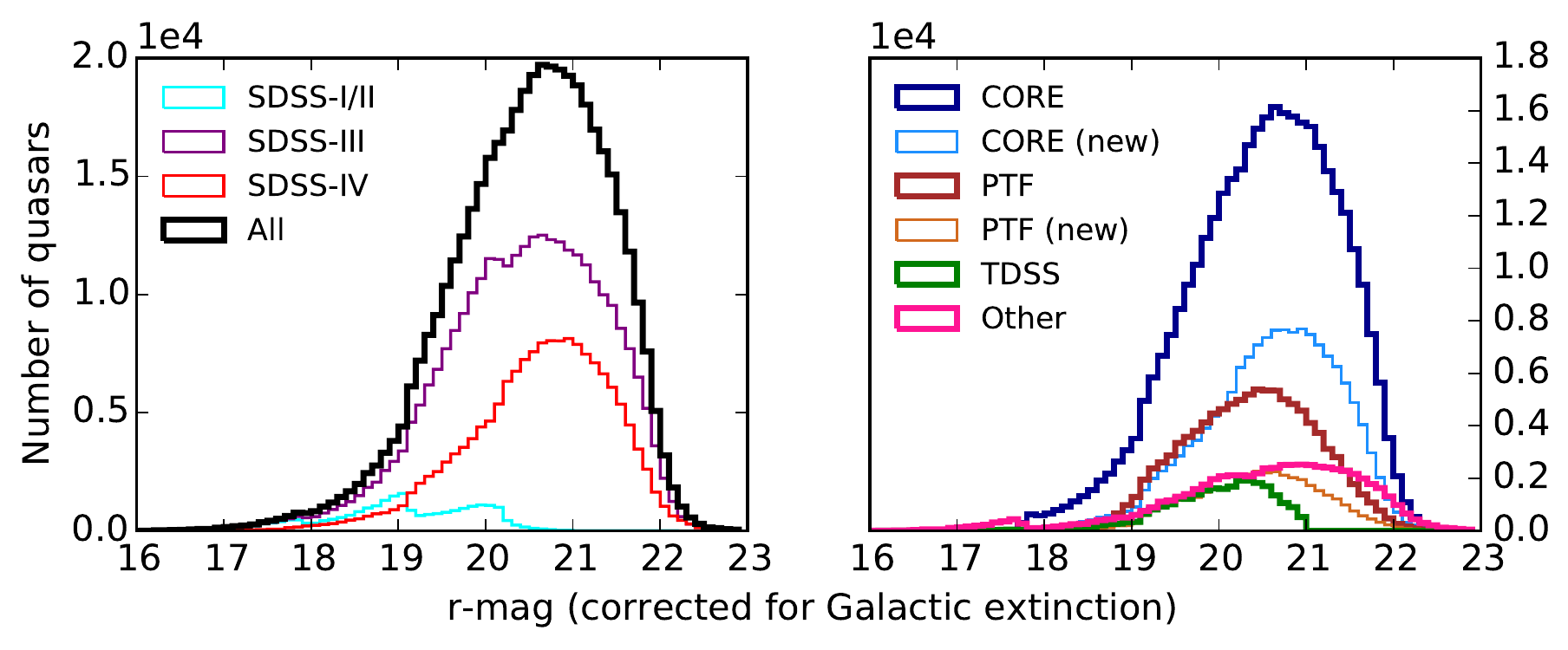}}
\caption{ 
\textit{Left Panel: }
Distribution of $r$-band magnitude corrected for Galactic extinction using the \cite{schlafly2011} dust maps for all DR14Q quasars (thick black histogram), quasars observed during SDSS-I/II (cyan histogram), SDSS-III (purple histogram) and SDSS-IV (red histogram).
\textit{Right Panel: }
Distribution of $r$-band magnitude corrected for Galactic extinction for all {\tt CORE} quasars (dark blue histogram), {\tt CORE} quasars observed as part of SDSS-IV only (light blue histogram), all {\tt PTF} quasars (brown histogram), {\tt PTF} quasars observed as part of SDSS-IV only (orange histogram), SDSS-IV/TDSS quasars (green histogram), and quasars selected as part of ancillary programs (pink histogram).
A given quasar can be selected by several target selection algorithms, hence the cumulative number of quasars in a $r$-band magnitude bin can exceed the total number of objects in it.
The bin size for both panels is $\Delta r = 0.1$.
}
\label{fig:rmag_distri}
\end{figure*}

A similar comparison is done for the Galactic-extinction corrected $r$-band magnitudes of DR14Q quasars in \Fig{rmag_distri} using the same color code as in \Fig{zdistri}.
The left panel of \Fig{rmag_distri} shows the $r$-band magnitude (corrected for Galactic extinction) distributions of quasars observed by each iteration of SDSS.
The right panel of \Fig{rmag_distri} displays the $r$-band magnitude distribution for each of the subsamples ({\tt CORE}, variability-selected quasars, TDSS and ancillary programs). 

\begin{figure}[htbp]
	\centering{\includegraphics[width=\linewidth]{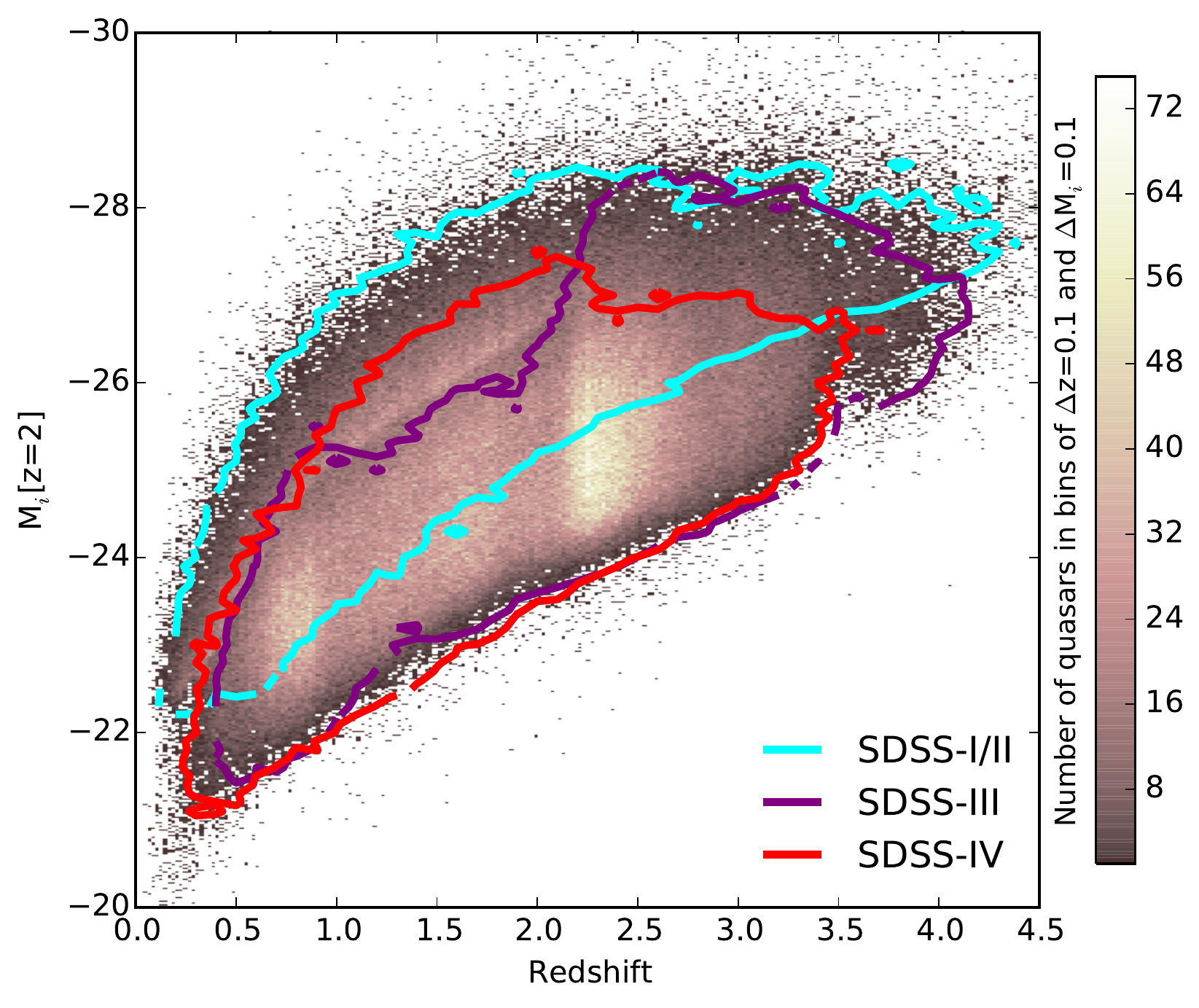}}
\caption{ 
Density map of the DR14Q quasars in the $L-z$ plane. The color map indicates the number of DR14Q quasars in bins of $\Delta z = 0.1$ and $\Delta {\rm M _i \left[ z=2 \right]} = 0.1$.
Colored contours correspond to the envelope in the $L-z$ plane for each iteration of SDSS. The absolute magnitudes assume ${\rm H_0 = 70 \ km \ s ^{-1} \ Mpc ^{-1}}$ and the K-correction is given by \cite{richards2006}, who define $K \left( z = 2 \right) = 0$.
}
\label{fig:absmag_redshift}
\end{figure}

Finally, we present a density map of the DR14Q quasars in the $L-z$ plane in \Fig{absmag_redshift}.
The area covered in this plane by each phase of SDSS is also displayed.
SDSS-I/II (cyan contour) has observed the brighest quasars at all redshifts.
SDSS-III (purple contour) has observed up to two magnitudes deeper than SDSS-I/II, mostly at $z > 2$.
SDSS-IV (red contour) is observing at the same depth as for SDSS-III but at lower redshift, i.e. focusing on the redshift range 0.8--2.2.

\section{Multiwavelength cross-correlation}
\label{s:multilambda}

We provide multi-wavelength matching of DR14Q quasars to several surveys:    
the FIRST radio survey \citep{becker1995}, the Galaxy Evolution Explorer 
\citep[GALEX,][]{martin2005} survey in the UV, the Two Micron All Sky Survey \citep[2MASS,][]{cutri2003,skrutskie2006},  the UKIRT Infrared Deep 
Sky Survey \citep[UKIDSS;][]{lawrence2007}, 
the Wide-Field Infrared Survey \citep[WISE,][]{wright2010}, the ROSAT All-Sky Survey \citep[RASS;][]{voges1999,voges2000}, and the 
seventh data release of the Third XMM-Newton Serendipitous
Source Catalog \citep{rosen2016}.

\subsection{FIRST}
\label{s:first}

As for the previous SDSS-III/BOSS quasar catalogs, we matched the DR14Q quasars
to the latest FIRST catalog \citep[December 2014; ][]{becker1995} using a 
2\arcsec \ matching radius.
We report the flux peak density at 20 cm and the signal-to-noise ratio of the detection.
Among the DR14Q quasars, \numprint{73126} lie outside of the FIRST footprint and have their {\tt FIRST\_MATCHED} flag set to -1.

A total of \numprint{18273} quasars have FIRST counterparts in DR14Q.
We estimate the fraction of chance superpositions by offsetting the declination of DR14Q quasars by 200\arcsec . We then re-match to the FIRST source catalog. We conclude that there are about 0.2\% of false positives in the DR14Q-FIRST matching.

\subsection{GALEX}
\label{s:galex}

As for DR12Q, GALEX \citep{martin2005} images are force-photometered (from GALEX Data Release 5) at the SDSS-DR8 centroids \citep{DR8}, such that low S/N point-spread function fluxes of objects not detected by GALEX are recovered,  
for both the FUV (\numprint{1350}-\numprint{1750} \AA ) and NUV (\numprint{1750}-\numprint{2750} \AA ) bands when available.
A total of \numprint{382838} quasars are detected in the NUV band, \numprint{304705} in the FUV band and \numprint{515728} have non-zero fluxes in both bands.

\subsection{2MASS}
\label{s:2mass}

We cross-correlate DR14Q with the All-Sky Data Release Point Source catalog \citep{skrutskie2006} using a matching radius of 2\arcsec.
We report the Vega-based magnitudes in the J, H and K-bands and their error together with the signal-to-noise ratio of the detections.
We also provide the value of the 2MASS flag rd\_flg[1], which defines the peculiar values of the magnitude and its error for each band\footnote{see http://www.ipac.caltech.edu/2mass/releases/allsky/doc/explsup.html for more details}.

There are \numprint{16427} matches in the catalog. This number is quite small compared with the number of DR14Q quasars because the sensitivity of 2MASS is much less than that of SDSS.
Applying the same method as described in \Sec{first}, we estimate that 0.8\% of the matches are false positives.

\subsection{WISE}
\label{s:wise}

We matched the DR14Q to the AllWISE Source Catalog\footnote{http://wise2.ipac.caltech.edu/docs/release/allwise/} \citep{wright2010,mainzer2011}.
Our procedure is the same as in DR12Q, with a matching radius of 2.0\arcsec .
There are \numprint{401980} matches from the AllWISE Source Catalog.
Following the procedure described in \Sec{first}, we estimate the rate of false positive matches to be about 2\%, which is consistent with the findings of \cite{krawczyk2013}.

We report the magnitudes, their associated errors, the S/N of the detection and reduced $\chi ^2$ of the profile-fitting in 
the four WISE bands centered at wavelengths of 3.4, 4.6, 12 and 22~$\mu$m. 
These magnitudes are in the Vega system, and are measured with profile-fitting photometry.
We also report the WISE catalog contamination and confusion flag, {\tt cc\_flags}, and their photometric quality flag, {\tt ph\_qual}.
As suggested on the WISE ``Cautionary Notes" page\footnote{http://wise2.ipac.caltech.edu/docs/release/allsky/expsup/sec1\_4b.html \#unreliab}, we
recommend using only those matches with {\tt cc\_flags} = ``0000'' to exclude objects that are flagged as spurious detections of image artifacts in any band.
Full details about quantities provided in the AllWISE Source Catalog can be found on their online documentation\footnote{http://wise2.ipac.caltech.edu/docs/release/allsky/expsup/ sec2\_2a.html}.

\subsection{UKIDSS}
\label{s:ukidss}

As for DR12Q, near infrared images from the UKIRT Infrared Deep Sky Survey \citep[UKIDSS; ][]{lawrence2007} are force-photometered.

We provide the fluxes and their associated errors, expressed in ${\rm W \ m^{-2} \ Hz^{-1}}$, in the Y, J, H and  K bands.
The conversion to the Vega magnitudes, as used in 2MASS, is given by the formula:
 \begin{equation}
 	{\rm mag _{X}} = -2.5 \times \log \frac{f_{\rm X}}{f_{\rm 0, X} \times 10^{-26}},
 \end{equation}
where X denotes the filter and the zero-point values $f_{\rm 0, X}$ are 2026, 1530, 1019 and 631 for the Y, J, H and K bands respectively.

A total of \numprint{112012} quasars are detected in at least one of the four bands Y, J, H or K.
\numprint{111083} objects are detected in the Y band, \numprint{110691} in the J band, \numprint{110630} in H band, \numprint{111245} in the K band and \numprint{108392} objects have 
non-zero fluxes in the four bands. Objects with zero fluxes lie outside the UKIDSS footprint.
The UKIDSS limiting magnitude is ${\rm K} \sim 18$ (for the Large Area Survey) while the 2MASS limiting magnitude in the same band is $\sim 15.3$.
This difference in depth between the two surveys explains the large difference in the numbers of matches with DR14Q.

\subsection{ROSAT}
\label{s:rosat}

As was done for the previous SDSS-III/BOSS quasar catalogs, we matched the DR14Q quasars to the ROSAT all sky survey Faint \citep{voges2000} and Bright \citep{voges1999} source 
catalogues with a matching radius of 30\arcsec.
Only the most reliable detections are included in our catalog: when the quality detection is flagged as potentially problematic, we do not include 
the match.
A total of \numprint{8655} quasars are detected in one of the RASS catalogs.
As for the cross-correlations described above, we estimate that 2.1\% of the RASS-DR14Q matches are due to chance superposition.

\subsection{XMM-Newton}
\label{s:xmm}

DR14Q was cross-correlated with the seventh data release of the Third XMM-Newton Serendipitous
Source Catalog \citep{rosen2016}\footnote{http://xmmssc.irap.omp.eu/Catalogue/3XMM-DR7/3XMM\_DR7.html} (3XMM-DR7) using a standard 5.0\arcsec \ matching radius.
For each of the \numprint{14736} DR14Q quasars with XMM-Newton counterparts, 
we report the soft (0.2--2 keV), hard (4.5--12 keV) and total (0.2--12 keV) fluxes, and associated errors, that were
computed as the weighted average of all the detections in the three XMM-Newton cameras (MOS1, MOS2, PN).
Corresponding observed X-ray luminosities are computed in each band and are not absorption corrected.
All fluxes and errors are expressed in ${\rm erg \ cm^{-2} \ s^{-1}}$ and luminosities are computed using the redshift value {\tt Z} from the present catalog.

\section{Description of the DR14Q catalog}
\label{s:description}

The DR14Q catalog is publicly available on the SDSS public website\footnote{http://www.sdss.org/dr14/algorithms/qso\_catalog} as a binary FITS table file.
All the required documentation (format, name, unit for each column) is provided in the FITS header. It is also summarized in \Tab{DR14Qformat}.
\\

\begin{longtab}
\begin{longtable}{clcl}
\caption{\label{t:DR14Qformat} Format of the binary FITS table containing DR14Q.}\\
\hline\hline
Column & Name & Format & Description \\
\hline
\endfirsthead
\caption{continued.}\\
\hline\hline
Column & Name & Format & Description \\
\hline
\endhead
\hline
\endfoot
1    & SDSS\_NAME           & STRING       & SDSS-DR14 designation -- hhmmss.ss+ddmmss.s (J2000)\\
2    & RA                   & DOUBLE       & Right Ascension in decimal degrees (J2000) \\ 
3    & DEC                  & DOUBLE       & Declination in decimal degrees (J2000) \\
4    & THING\_ID            & INT32        & Unique SDSS source identifier \\
5    & PLATE                & INT32        & Spectroscopic plate number \\
6    & MJD                  & INT32        & Modified Julian Day of the spectroscopic \\
     &                      &              & observation \\
7    & FIBERID              & INT32        & Spectroscopic fiber number \\
8    & SPECTRO              & STRING       & SDSS or BOSS \\
\hline
9    & Z                    & DOUBLE       & Redshift (most robust estimate for each quasar) \\
10   & Z\_ERR               & DOUBLE       & Error on redshift given in Col. \#9 \\
11   & SOURCE\_Z            & STRING       & Origin of the redshift measurement in Col. \#9 \\
     &                      &              & (VI, PIPE, AUTO, OTHER) \\
12   & Z\_VI                & DOUBLE       & Redshift based on visual inspection (when \\
     &                      &              & available) \\
13   & Z\_PIPE              & DOUBLE       & SDSS pipeline redshift estimate \\
14   & Z\_PIPE\_ERR         & DOUBLE       & Error on SDSS pipeline redshift estimate \\
15   & ZWARNING             & INT32        & Quality flag on SDSS pipeline redshift estimate \\
16   & Z\_PCA               & DOUBLE       & PCA redshift (homogeneous over the full sample) \\
17   & Z\_PCA\_ERR          & DOUBLE       & Error on PCA redshift \\
18   & Z\_MGII              & DOUBLE       & Redshift of the \ion{Mg}{ii} emission line \\
\hline
19   & BOSS\_TARGET1        & INT64        & BOSS target selection flag for main survey \\
20   & ANCILLARY\_TARGET1   & INT64        & BOSS target selection flag for ancillary programs \\
21   & ANCILLARY\_TARGET2   & INT64        & BOSS target selection flag for ancillary programs \\
22   & EBOSS\_TARGET0       & INT64        & Target selection flag for the eBOSS pilot survey \\
23   & EBOSS\_TARGET1       & INT64        & eBOSS target selection flag \\
24   & EBOSS\_TARGET2       & INT64        & eBOSS target selection flag \\
\hline
25   & N\_SPEC\_SDSS        & INT32        & Number of additional SDSS spectra \\
26   & N\_SPEC\_BOSS        & INT32        & Number of additional BOSS spectra \\
27   & N\_SPEC              & INT32        & Number of additional spectra (SDSS and BOSS)\\
28   & PLATE\_DUPLICATE     & INT32[51]    & Spectroscopic plate number for each duplicate \\
     &                      &              & spectrum \\
29   & MJD\_DUPLICATE       & INT32[51]    & Spectroscopic MJD for each duplicate spectrum \\
30   & FIBERID\_DUPLICATE   & INT32[51]    & Fiber number for each duplicate spectrum\\
31   & SPECTRO\_DUPLICATE   & STRING[51]   & SDSS or BOSS for each duplicate spectrum \\
\hline
\hline
32   & BI\_CIV              & FLOAT        & BALnicity Index of \ion{C}{iv} absorption trough (in ${\rm km \ s^{-1}}$) \\
33   & ERR\_BI\_CIV         & FLOAT        & Error on the BALnicity index of \ion{C}{iv} trough (in ${\rm km \ s^{-1}}$) \\
\hline
\hline
%
%
34   & RUN\_NUMBER          & INT32         & SDSS Imaging Run Number of photometric measurements \\
35   & RERUN\_NUMBER        & STRING        & SDSS Photometric Processing Rerun Number \\
36   & COL\_NUMBER          & INT32         & SDSS Camera Column Number (1-6) \\
37   & FIELD\_NUMBER        & INT32         & SDSS Field Number \\
38   & OBJ\_ID              & STRING        & SDSS Object Identification Number \\
39   & PSFFLUX              & FLOAT[5]     & Flux in the $u$, $g$, $r$, $i$ and $z$ bands \\
40   & IVAR\_IVAR           & FLOAT[5]     & Inverse variance of the $u$, $g$, $r$, $i$ and $z$ fluxes \\
41   & PSFMAG               & FLOAT[5]     & PSF magnitudes in $u$, $g$, $r$, $i$ and $z$ bands \\
42   & ERR\_PSFMAG          & FLOAT[5]     & Error on PSF magnitudes \\
43  & MI                   & FLOAT        & Absolute magnitude in $i$ band $M_i \left[ {\rm z = 2} \right]$ with  \\
     &                      &              & $H_0 = 70 \ {\rm km \ s^{-1} \ Mpc^{-1}, \ } \Omega_M = 0.3, \ \Omega_{\Lambda} = 0.7,$ \\
     &                      &              &  $\alpha _{\nu} = -0.5$ \\
44  & GAL\_EXT             & FLOAT[5]     & Galactic extinction in the five SDSS bands \\
     &                      &              & (Schlafly et al., 2011)\\
\hline
\hline
%
45   & RASS\_COUNTS                    &  DOUBLE       & log RASS full band count rate (counts s$^{-1}$)\\
46   & RASS\_COUNTS\_SNR               &  FLOAT        & S/N of the RASS count rate \\
47   & SDSS2ROSAT\_SEP                 &  DOUBLE       & SDSS-RASS separation in arcsec \\
48   & FLUX\_0.2\_2.0keV               & FLOAT         & Soft (0.2--2.0 keV) X-ray flux from XMM-Newton \\
     &                                 &               & expressed in ${\rm erg \ s^{-1} \ cm^{-2}}$ \\
49   & FLUX\_0.2\_2.0keV\_ERR          & FLOAT         & Error on soft X-ray flux from XMM-Newton \\
     &                                 &               & (in ${\rm erg \ s^{-1} \ cm^{-2}}$) \\
50   & FLUX\_2.0\_12.0keV              & FLOAT         & Hard (4.5--12.0 keV) X-ray flux from XMM-Newton \\
     &                                 &               & expressed in ${\rm erg \ s^{-1} \ cm^{-2}}$ \\
51   & FLUX\_2.0\_12.0keV\_ERR         & FLOAT         & Error on hard X-ray flux from XMM-Newton \\
     &                                 &               & (in ${\rm erg \ s^{-1} \ cm^{-2}}$) \\
52   & FLUX\_0.2\_12.0keV              & FLOAT         & Total (0.2--12.0 keV) X-ray flux from XMM-Newton \\
     &                                 &               & expressed in ${\rm erg \ s^{-1} \ cm^{-2}}$ \\
53   & FLUX\_0.2\_12.0keV\_ERR         & FLOAT         & Error on total X-ray flux from XMM-Newton \\
     &                                 &               & (in ${\rm erg \ s^{-1} \ cm^{-2}}$) \\
54   & LUM\_0.2\_12.0keV               & FLOAT         & Total (0.2--12.0 keV) X-ray luminosity from XMM-Newton \\
     &                                 &               & expressed in ${\rm erg \ s^{-1}}$ \\
55   & SDSS2XMM\_SEP                   & DOUBLE        & SDSS-XMM-Newton separation in arcsec \\
56   & GALEX\_MATCHED                  & SHORT         & GALEX match flag  \\
57   & FUV                             & DOUBLE        & $fuv$ flux (GALEX) \\
58   & FUV\_IVAR                       & DOUBLE        & Inverse variance of $fuv$ flux \\
59   & NUV                             & DOUBLE        & $nuv$ flux (GALEX) \\
60   & NUV\_IVAR                       & DOUBLE        & Inverse variance of $nuv$ flux \\
61   & JMAG                            & DOUBLE        & $J$ magnitude (Vega, 2MASS) \\
62   & ERR\_JMAG                       & DOUBLE        & Error in $J$ magnitude \\
63   & JSNR                            & DOUBLE        & J-band S/N \\
64   & JRDFLAG                         & INT32         & J-band photometry flag\\ 
65   & HMAG                            & DOUBLE        & $H$ magnitude (Vega, 2MASS) \\
66   & ERR\_HMAG                       & DOUBLE        & Error in $H$ magnitude \\
67   & HSNR                            & DOUBLE        & H-band S/N \\
68   & HRDFLAG                         & INT32         & H-band photometry flag\\ 
69   & KMAG                            & DOUBLE        & $K$ magnitude (Vega, 2MASS) \\
70   & ERR\_KMAG                       & DOUBLE        & Error in $K$ magnitude \\
71   & KSNR                            & DOUBLE        & K-band S/N \\
72   & KRDFLAG                         & INT32         & K-band photometry flag\\ 
73   & SDSS2MASS\_SEP                  & DOUBLE        & SDSS-2MASS separation in arcsec \\
74   & W1MAG                           & DOUBLE        & $w1$ magnitude (Vega, WISE) \\
75   & ERR\_W1MAG                      & DOUBLE        & Error in $w1$ magnitude \\
76   & W1SNR                           & DOUBLE        & S/N in w1 band    \\
77   & W1CHI2                          & DOUBLE        & $\chi^2$ in w1 band \\
78   & W2MAG                           & DOUBLE        & $w2$ magnitude (Vega, WISE) \\
79   & ERR\_W2MAG                      & DOUBLE        & Error in $w2$ magnitude \\
80   & W2SNR                           & DOUBLE        & S/N in w2 band    \\
81   & W2CHI2                          & DOUBLE        & $\chi^2$ in w2 band \\
82   & W3MAG                           & DOUBLE        & $w3$ magnitude (Vega, WISE) \\
83   & ERR\_W3MAG                      & DOUBLE        & Error in $w3$ magnitude \\
84   & W3SNR                           & DOUBLE        & S/N in w3 band    \\
85   & W3CHI2                          & DOUBLE        & $\chi^2$ in w3 band \\
86   & W4MAG                           & DOUBLE        & $w4$ magnitude  (Vega, WISE) \\
87   & ERR\_W4MAG                      & DOUBLE        & Error in $w4$ magnitude \\
88   & W4SNR                           & DOUBLE        & S/N in w4 band    \\
89   & W4CHI2                          & DOUBLE        & $\chi^2$ in w4 band \\
90   & CC\_FLAGS                       & STRING        & WISE contamination and confusion flag \\
91   & PH\_FLAG                        & STRING        & WISE photometric quality flag \\
92   & SDSS2WISE\_SEP                  & DOUBLE        & SDSS-WISE separation in arcsec \\
93   & UKIDSS\_MATCHED                 & SHORT         & UKIDSS Matched \\
94   & YFLUX                           & DOUBLE        & Y-band flux density from UKIDSS (in ${\rm W \ m^{-2} \ Hz^{-1}}$) \\
95   & YFLUX\_ERR                      & DOUBLE        & Error in Y-band density flux from UKIDSS (in ${\rm W \ m^{-2} \ Hz^{-1}}$) \\
96   & JFLUX                           & DOUBLE        & J-band flux density from UKIDSS (in ${\rm W \ m^{-2} \ Hz^{-1}}$)\\
97   & JFLUX\_ERR                      & DOUBLE        & Error in J-band flux density from UKIDSS (in ${\rm W \ m^{-2} \ Hz^{-1}}$) \\
98   & HFLUX                           & DOUBLE        & H-band flux density from UKIDSS (in ${\rm W \ m^{-2} \ Hz^{-1}}$)\\
99   & HFLUX\_ERR                      & DOUBLE        & Error in H-band flux density from UKIDSS (in ${\rm W \ m^{-2} \ Hz^{-1}}$)\\
100  & KFLUX                           & DOUBLE        & K-band flux density from UKIDSS (in ${\rm W \ m^{-2} \ Hz^{-1}}$)\\
101  & KFLUX\_ERR                      & DOUBLE        & Error in K-band flux density from UKIDSS  (in ${\rm W \ m^{-2} \ Hz^{-1}}$) \\
102  & FIRST\_MATCHED                  & INT           & FIRST matched \\
103  & FIRST\_FLUX                     & DOUBLE        & FIRST peak flux density at 20 cm expressed in mJy \\
104  & FIRST\_SNR                      & DOUBLE        & S/N of the FIRST flux density \\
105  & SDSS2FIRST\_SEP                 & DOUBLE        & SDSS-FIRST separation in arcsec \\
\end{longtable}
\end{longtab}

\par\noindent
Notes on the catalog columns:\\
\smallskip
\noindent
{\bf 1}. The DR14 object designation, given by the format \hbox{SDSS Jhhmmss.ss+ddmmss.s}; only the final 18
characters  are listed in the catalog (i.e., the character string \hbox{"SDSS J"}  is dropped).
The coordinates in the object name follow IAU convention and are truncated, not rounded.

\noindent
{\bf 2-3}. The J2000 coordinates (Right Ascension and Declination) in decimal degrees.  
The astrometry is from SDSS-DR14 \citep{DR14}.

\noindent
{\bf 4}.  The 64-bit integer that uniquely describes the objects
that are listed in the SDSS (photometric and spectroscopic) catalogs ({\tt THING\_ID}).

\noindent
{\bf 5-7}. Information about the spectroscopic observation (Spectroscopic plate number, 
Modified Julian Date, and spectroscopic fiber number) used to
determine the characteristics of the spectrum.
These three numbers are unique for each spectrum, and
can be used to retrieve the digital spectra from the public SDSS database.
When an object has been observed more than once, we selected the best quality spectrum as 
defined by the SDSS pipeline \citep{bolton2012}, i.e. with {\tt SPECPRIMARY}~=~1.

\noindent
{\bf 8}. DR14Q compiles all spectroscopic observations of quasars, including SDSS-I/II spectra taken with a different spectrograph. For spectra taken with the SDSS spectrographs, i.e. spectra released prior to SDSS-DR8 \citep{DR8}, {\tt SPECTRO} is set to ''SDSS''. For spectra taken with the BOSS spectrographs \citep{smee2013}, {\tt SPECTRO} is set to ''BOSS''.

\noindent
{\bf 9-11}. Quasar redshift (col. \#9) and associated error (col. \#10). This redshift estimate is the most robust for quasar cataloging purposes and it is used as a prior for refined redshift measurements. 
Values reported in col. \#9 are from different sources: the outcome of the automated procedure described in \Sec{def_autoclass}, visual inspection or the BOSS pipeline \citep{bolton2012}. The origin of the redshift value is given in col. \#11 (AUTO, VI and PIPE respectively).

\noindent
{\bf 12}. Redshift from the visual inspection, {\tt Z\_VI}, when available. All SDSS-I/II/III quasars have been visually inspected. About 7\% of SDSS-IV quasars have been through this process (see \Sec{def_autoclass} for more details).

\noindent
{\bf 13-15}. Redshift ({\tt Z\_PIPE}, col. \#13), associated error ({\tt Z\_PIPE\_ERR}, col. \#14) and quality flag ({\tt ZWARNING}, col. \#15) from the BOSS pipeline \citep{bolton2012}. ZWARNING~$>$~0 indicates uncertain results in the redshift-fitting code. 

\noindent
{\bf 16-17}.  Automatic redshift estimate ({\tt Z\_PCA}, col. \#16) and associated error ({\tt Z\_PCA\_ERR}, col. \#17) using a linear combination of four principal components (see \Sec{redshift} for details). 
When the velocity difference between the automatic PCA and visual inspection redshift estimates is larger than 
\numprint{5000}~${\rm km \ s^{-1}}$, this PCA redshift and error are set to $-1$.

\noindent
{\bf 18}. Redshifts measured from the \MgII\ emission line from a linear combination of five principal components \citep[see ][]{paris2012}. The line redshift is estimated using the position of the maximum of each emission line, contrary to {\tt Z\_PCA} (column \#16) which is a global estimate using all the information available in a given spectrum.

\noindent
{\bf 19-24}.
The main target selection information for SDSS-III/BOSS quasars is tracked with the {\tt BOSS\_TARGET1} flag bits \citep[col \#19; see Table 2 in ][ for a full description]{ross2012}.
SDSS-III ancillary program target selection is tracked 
with the {\tt ANCILLARY\_TARGET1} (col. \#20) and {\tt ANCILLARY\_TARGET2} (col. \#21) flag bits.
The bit values and the corresponding program names are listed in \cite{dawson2013}, and \cite{DR12}. 
Target selection information for the SDSS-IV pilot survey \citep[SEQUELS; ][]{dawson2016,myers2015} is tracked with the {\tt EBOSS\_TARGET0} flag bits (col. \#22).
Finally, target selection information for SDSS-IV/eBOSS, SDSS-IV/TDSS and SDSS-IV/SPIDERS quasars is tracked with the {\tt EBOSS\_TARGET1} and {\tt EBOSS\_TARGET2} flag bits.
All the target selection bits, program names and associated references are summarized in \Tab{qts_bit}.

\noindent
{\bf 25-31}. If a quasar in DR14Q was observed more than once by SDSS-I/II/III/IV, the number of additional SDSS-I/II spectra is given by {\tt N\_SPEC\_SDSS} (col. \#25), the number of additional SDSS-III/IV spectra by {\tt N\_SPEC\_BOSS} (col. \#26), and the total number by {\tt N\_SPEC} (col. \#27).
The associated plate ({\tt PLATE\_DUPLICATE}, MJD ({\tt MJD\_DUPLICATE}), fiber ({\tt FIBERID\_DUPLICATE}) numbers, and spectrograph information ({\tt SPECTRO\_DUPLICATE}) are given in Col. \#28, 29, 30 and 31 respectively.
If a quasar was observed N times in total, the best spectrum is identified in Col. \#5-7, the corresponding {\tt N\_SPEC} is N-1, and the first N-1 columns of {\tt PLATE\_DUPLICATE},
{\tt MJD\_DUPLICATE}, {\tt FIBERID\_DUPLICATE}, and {\tt SPECTRO\_DUPLICATE} are filled with relevant information. Remaining columns are set to -1.

\noindent
{\bf 32-33}. Balnicity index \citep[BI; ][; col. \#32]{weymann1991} for \CIV\ troughs, and associated error (col. \#33), expressed in ${\rm km \ s^{-1}}$. See definition in \Sec{bal}.
The Balnicity index is measured for quasars with $z > 1.57$ only, so that the trough enters into the BOSS wavelength region.
In cases with poor fits to the continuum, the balnicity index and its error are set to $-1$.

\noindent
{\bf 34}. The SDSS Imaging Run number ({\tt RUN\_NUMBER}) of the photometric observation used in the catalog.

\noindent
{\bf 35-38}. Additional SDSS processing information: the photometric processing rerun number ({\tt RERUN\_NUMBER}, col. \#35); the camera column (1--6) containing the image of the object ({\tt COL\_NUMBER}, col. \#36), the field number of the run containing the object ({\tt FIELD\_NUMBER}, col. \#37),
and the object identification number
\citep[{\tt OBJ\_ID}, col. \#38; see][for descriptions of these parameters]{stoughton2002}.

\noindent
{\bf 39-40}. DR14 PSF fluxes, expressed in nanomaggies\footnote{See http://www.sdss.org/dr14/algorithms/magnitudes/\#nmgy}, and inverse variances (not corrected for Galactic extinction) in the five SDSS filters. 

\noindent
{\bf 41-42}. DR14 PSF AB magnitudes \citep{oke1983} and errors (not corrected for Galactic extinction) in the five SDSS filters.
These magnitudes are Asinh magnitudes as defined in \cite{lupton1999}.

\noindent
{\bf 43}. The absolute magnitude in the $i$ band at $z=2$ calculated 
using a power-law (frequency)
continuum index of~$-0.5$.
The K-correction is computed using Table~4 from \cite{richards2006}.
We use the SDSS primary photometry to compute this value.

\noindent
{\bf 44}. Galactic extinction in the five SDSS bands based on \cite{schlafly2011}.

\noindent
{\bf 45}. The logarithm of the vignetting-corrected count rate (photons s$^{-1}$)
in the broad energy band \hbox{(0.1--2.4 keV)} from the
{\it ROSAT} All-Sky Survey Faint Source Catalog \citep{voges2000} and the
{\it ROSAT} All-Sky Survey Bright Source Catalog \citep{voges1999}.
The matching radius was set to 30\arcsec \ (see \Sec{rosat}).

\noindent
{\bf 46}. The S/N of the {\it ROSAT} measurement.

\noindent
{\bf 47}. Angular separation between the SDSS and {\it ROSAT} All-Sky Survey
locations (in arcseconds).

\noindent
{\bf 48-49}. Soft X-ray flux (0.2--2 keV) from XMM-Newton matching, expressed in ${\rm erg \ cm^{-2} \ s^{-1}}$, and its error. 
In the case of multiple observations, the values reported here are the weighted average of all the XMM-Newton detections in this band.

\noindent
{\bf 50-51}. Hard X-ray flux (4.5--12 keV) from XMM-Newton matching, expressed in ${\rm erg \ cm^{-2} \ s^{-1}}$, and its error.
In the case of multiple observations, the reported values are the weighted average of all the XMM-Newton detections in this band.

\noindent
{\bf 52-53}. Total X-ray flux (0.2--12 keV) from the three XMM-Newton CCDs (MOS1, MOS2 and PN), expressed in ${\rm erg \ cm^{-2} \ s^{-1}}$, and its error. In the case of multiple XMM-Newton observations, only the longest
exposure was used to compute the reported flux.

\noindent
{\bf 54}. Total X-ray luminosity (0.2--12 keV) derived from the flux computed in Col. \#52, expressed in ${\rm erg \ s^{-1}}$. 
This value is computed using the redshift value reported in col. \#9 and is not absorption corrected.

\noindent
{\bf 55}. Angular separation between the XMM-Newton and SDSS-DR14 locations, expressed in arcsec.

\noindent
{\bf 56}. If a SDSS-DR14 quasar matches with GALEX photometring, {\tt GALEX\_MATCHED} is set to 1, 0 if no GALEX match.

\noindent
{\bf 57-60}. UV fluxes and inverse variances from GALEX, aperture-photometered from the original GALEX images in the two bands FUV and NUV. The fluxes are expressed in nanomaggies.

\noindent
{\bf 61-62}. The $J$ magnitude and error from the Two Micron All Sky Survey
All-Sky Data Release Point Source Catalog \citep{cutri2003} using
a matching radius of ~2.0\arcsec\ (see \Sec{2mass}).  A non-detection by 2MASS is indicated by a "0.000" in these columns.  
The 2MASS measurements are in Vega, not AB, magnitudes.  

\noindent
{\bf 63-64}. Signal-to-noise ratio in the $J$ band and corresponding 2MASS {\tt jr\_d} flag that gives the meaning of the peculiar values of the magnitude and its error\footnote{see http://www.ipac.caltech.edu/2mass/releases/allsky/doc/ explsup.html}.

\noindent
{\bf 65-68}. Same as 61-64 for the $H$-band.

\noindent
{\bf 69-72}. Same as 61-64 for the $K$-band.

\noindent
{\bf 73}. Angular separation between the SDSS-DR14 and 2MASS positions (in arcsec).

\noindent
{\bf 74-75}. The $w1$  magnitude and error from the Wide-field Infrared Survey Explorer
\citep[WISE;][]{wright2010} AllWISE Data Release Point Source Catalog  using a matching radius of 2\arcsec.

\noindent
{\bf 76-77}. Signal-to-noise ratio and $\chi^2$ in the WISE $w1$ band.

\noindent
{\bf 78-81}. Same as 74-77 for the $w2$-band.

\noindent
{\bf 82-85}. Same as 74-77 for the $w3$-band.

\noindent
{\bf 86-89}. Same as 74-77 for the $w4$-band.

\noindent
{\bf 90}. WISE contamination and confusion flag.

\noindent
{\bf 91}. WISE photometric quality flag.

\noindent
{\bf 92}. Angular separation between SDSS-DR14 and WISE positions (in arcsec).

\noindent
{\bf 93}. If a SDSS-DR14 quasar matches UKIDSS aperture-photometering data, {\tt UKIDSS\_MATCHED} is set to 1, it is set to 0 if UKIDSS match.

\noindent
{\bf 94-101}. Flux density and error from UKIDSS, aperture-photometered from the original UKIDSS images in the four bands Y (Col. \#94-95), J (Col. \#96-97), 
H (Col. \#98-99) and K (Col. \#100-101). The fluxes and errors are expressed in ${\rm W \ m^{-2} \ Hz^{-1}}$.

\noindent
{\bf 102}. If there is a source in the FIRST radio catalog (version December 2014) within 2.0\arcsec\
of the quasar position, the {\tt FIRST\_MATCHED} flag provided in this column is set to 1, 0 if not. If the quasar lies outside of the FIRST footprint, it is set to -1.

\noindent
{\bf 103}. The FIRST peak flux density, expressed in mJy.

\noindent
{\bf 104}. The signal-to-noise ratio of the FIRST source whose flux is given in Col. \#103.

\noindent
{\bf 105}. Angular separation between the SDSS-DR14 and FIRST positions (in arcsec).

\section{Conclusion}
\label{s:conclusion}

We  have  presented  the quasar  catalog  of  the  SDSS-IV/eBOSS survey 
corresponding to Data Release 14 of SDSS and resulting from the first two years of 
SDSS-IV observations. The catalog, DR14Q,  contains \numprint{526356}  quasars,
\numprint{144046} of which are new discoveries. We provide robust identification from 
the application of an automated procedure and partial visual inspection of about
10\% of the sample (likely ill-identified targets by the automated procedure).
Refined redshift measurements based  on  the  result  of  a  principal  component  
analysis  of  the spectra are also given. The present catalog contains about 
80\% more quasars at $z<2$ than our previous release \citep{paris2017}. 
As part of DR14Q, we  also  provide  a  catalog  of  \numprint{21877}  BAL  quasars  and  their
properties.  Multi-wavelength matching with GALEX, 2MASS, UKIDSS, WISE, FIRST, RASS and
XMM-Newton observations is also provided as part of DR14Q.
\par\noindent
The next SDSS public release containing new eBOSS data is scheduled for
the summer of 2019 and will contain spectroscopic data after four years of observations, which should represent more than \numprint{700000} quasars.

\begin{acknowledgements}

IP acknowledges the support of the OCEVU Labex (ANR-11-LABX-0060) and the A*MIDEX project (ANR-11-IDEX-0001-02) funded by the ``Investissements d’Avenir'' French government program managed by the ANR.
The French Participation Group to SDSS-IV was supported by the Agence Nationale de la Recherche under contracts ANR-16-CE31-0021.\\
AW acknowledges support from a Leverhulme Trust Early Career Fellowship.\\
ADM and BWL acknowledge support from National Science Foundation grants 1515404 and 1616168.\\
IP thanks warmly the Great and Extremely COol team, especially Samuel Boissier, Jean-Claude Bouret, Sylvain de la Torre, Audrey Delsanti, Olivier Groussin, Olivier Perfect Ilbert, \'Eric Jullo, Vincent Le Brun, and S\'ebastien Vives, for encouraging creativity and for dealing with its unexpected consequences.\\

This research has made use of data obtained from the 3XMM XMM-Newton serendipitous source catalogue compiled by the 10 institutes of the XMM-Newton Survey Science Centre selected by ESA.

      Funding for the Sloan Digital Sky Survey IV has been provided by
the Alfred P. Sloan Foundation, the U.S. Department of Energy Office of
Science, and the Participating Institutions. SDSS-IV acknowledges
support and resources from the Center for High-Performance Computing at
the University of Utah. The SDSS web site is www.sdss.org.

SDSS-IV is managed by the Astrophysical Research Consortium for the 
Participating Institutions of the SDSS Collaboration including the 
Brazilian Participation Group, the Carnegie Institution for Science, 
Carnegie Mellon University, the Chilean Participation Group, the French Participation Group, Harvard-Smithsonian Center for Astrophysics, 
Instituto de Astrof\'isica de Canarias, The Johns Hopkins University, 
Kavli Institute for the Physics and Mathematics of the Universe (IPMU) / 
University of Tokyo, Lawrence Berkeley National Laboratory, 
Leibniz Institut f\"ur Astrophysik Potsdam (AIP),  
Max-Planck-Institut f\"ur Astronomie (MPIA Heidelberg), 
Max-Planck-Institut f\"ur Astrophysik (MPA Garching), 
Max-Planck-Institut f\"ur Extraterrestrische Physik (MPE), 
National Astronomical Observatory of China, New Mexico State University, 
New York University, University of Notre Dame, 
Observat\'ario Nacional / MCTI, The Ohio State University, 
Pennsylvania State University, Shanghai Astronomical Observatory, 
United Kingdom Participation Group,
Universidad Nacional Aut\'onoma de M\'exico, University of Arizona, 
University of Colorado Boulder, University of Oxford, University of Portsmouth, 
University of Utah, University of Virginia, University of Washington, University of Wisconsin, 
Vanderbilt University, and Yale University.
\end{acknowledgements}

\bibliographystyle{aa}
\bibliography{DR14Q}

\end{document}